\documentclass[%
aps,%
 amsmath,amssymb,
 reprint,%
 superscriptaddress,
nofootinbib,
floatfix,
preprintnumbers
]{revtex4-1}

\usepackage{bbold}
\usepackage{graphicx}
\usepackage{dcolumn}
\usepackage{bm}
\usepackage[dvipsnames]{xcolor}
\graphicspath{{Images/}}

\usepackage[caption=false]{subfig}
\usepackage[normalem]{ulem}
\usepackage{braket}
\usepackage{esdiff}
\usepackage[hidelinks]{hyperref}
\usepackage{appendix}
\usepackage{mathtools}
\usepackage{amsmath}
\usepackage{calc}

\graphicspath{{Images/}}

\begin{document}

\preprint{SLAC-PUB-17657, N3AS-22-005}

\title{Collective neutrino oscillations with tensor networks\\using a time-dependent variational principle}

\author{Michael J.\ Cervia}
\email{cervia@gwu.edu}
\affiliation{%
Department of Physics, University of Wisconsin--Madison,
Madison, Wisconsin 53706, USA
}%
\affiliation{Department of Physics, The George Washington University,
Washington, District of Columbia 20052, USA
}
\affiliation{Department of Physics, University of Maryland, 
College Park, Maryland 20742, USA
}
\author{Pooja Siwach}
\email{psiwach@physics.wisc.edu}
\affiliation{%
Department of Physics, University of Wisconsin--Madison,
Madison, Wisconsin 53706, USA
}%
\author{Amol V.\ Patwardhan}
\email{apatward@slac.stanford.edu} 
\affiliation{%
SLAC National Accelerator Laboratory,
Menlo Park, California 94025, USA $\>\>\>\>\>\>$
}%
\author{A.\ B.\ Balantekin}
\email{baha@physics.wisc.edu}
\affiliation{%
Department of Physics, University of Wisconsin--Madison,
Madison, Wisconsin 53706, USA
}%
\author{S.\ N.\ Coppersmith}
\email{snc@physics.wisc.edu}
\affiliation{%
Department of Physics, University of Wisconsin--Madison,
Madison, Wisconsin 53706, USA
 }%
\affiliation{%
School of Physics, The University of New South Wales,
Sydney, New South Wales 2052,  Australia}
\author{Calvin W.\ Johnson}
\email{cjohnson@sdsu.edu}
\affiliation{
Department of Physics, San Diego State University, San Diego,
California 92182-1233, USA
}

\date{\today}

\begin{abstract}

If 
a system of flavor-oscillating neutrinos is at  high enough densities that neutrino-neutrino coherent forward scatterings are non-negligible, the system becomes a time-dependent many-body problem. 
An important and open question is whether the flavor evolution is sufficiently described by a mean-field approach or 
can be strongly affected by correlations arising from two-body interactions in the neutrino Hamiltonian, as measured by  nontrivial quantum entanglement. 
Numerical computations of the time evolution of many-body quantum systems are challenging because the size of the Hilbert space scales exponentially with the number of particles $N$ in the system. 
Thus, it is important to investigate approximate but beyond-mean-field numerical treatments 
at larger values of $N$. Here we investigate the efficacy of tensor network methods to calculate 
the time evolution of interacting neutrinos at larger values of $N$ than are possible with conventional methods. 
In particular, we introduce the use of time-dependent variational principle methods to address the {long-range} (in momentum space) interactions of the neutrino Hamiltonian when including many distinct vacuum oscillation frequencies. 
We also define new error measures based upon the instantaneously conserved charge operators known for this Hamiltonian to determine validity of large-$N$ tensor network calculations. 
\end{abstract}

\maketitle

\section{Introduction} \label{section:intro}

Collective effects in the flavor oscillations of neutrinos in environments where large fluxes of neutrinos are present, such as core collapse supernovae, neutron star mergers, or the early universe, have been a subject of great interest over the past few decades (e.g., \cite{Duan:2009cd,Duan:2010fr,Chakraborty:2016a,Tamborra:2020cul} and references therein). 
Such flavor oscillations of neutrinos could play an important role in the synthesis of elements in these environments, as well as in the supernova explosion mechanism itself~\cite{Fuller:1992eu,Qian:1993dg,Fuller:1993ry,Fuller:1995qy,Duan:2010af,Wu:2014kaa,Wu:2015glr,Sasaki:2017jry,Balantekin:2017bau,Xiong:2019nvw,Xiong:2020ntn}. Understanding  collective flavor oscillation effects is needed to  robustly interpret numerical simulations of these environments.

Despite the weakness of weak interactions, at sufficient density neutrino-neutrino interactions contribute substantially 
to the neutrino forward scattering potential, transforming collective neutrino oscillations into a quantum many-body problem.
As in any interacting many-body quantum system, the dimension of the Hilbert space describing the  wave function of the system grows exponentially with particle number. Consequently, the computational complexity grows exponentially as well, 
and it is untenable to fully solve the interacting many-neutrino system for the 
 large densities of neutrinos present in environments where collective effects matter.
 
To get around this roadblock, one frequently turns to \lq\lq mean-field\rq\rq\ treatments which neglect multineutrino quantum correlations~\cite{Pantaleone92,Pantaleone:1992eq,Sigl:1993fr,Qian95,McKellar:1994uq,Balantekin:2006tg}. Assessing the reliability of the mean-field approximation in this context has been a topic of long-standing interest, explored through the use of simplified models of interacting neutrinos~\cite{Bell:2003mg,Friedland:2003dv,Friedland:2003eh,Friedland:2006ke,Balantekin:2006tg,Pehlivan:2011hp,Volpe:2013lr,Pehlivan:2016lxx,Birol:2018qhx,Cervia:2019nzy,Patwardhan:2019zta,Rrapaj:2019pxz,Cervia:2019res,Colombi:2020egf,Roggero:2021asb,Roggero:2021fyo,Hall:2021rbv,Yeter-Aydeniz:2021olz,Patwardhan:2021rej,Xiong:2021evk,Martin:2021bri}. 
{Here in this paper, we conduct a }
comparison of advanced numerical techniques for time evolution of many-neutrino systems, and we {further explore} whether such time evolution brings about strong many-neutrino correlations, i.e., entanglement, 
{signaling a deviation} from mean-field approaches. We find that tensor network methods, described below, can provide a significant speedup, allowing 
us to reach much larger values of $N$ for certain initial conditions. {At these larger values of $N$,} our simulations {continue to find} significant entanglement in the many-neutrino system.

\subsection{Overview of past and present numerical approaches}

As in many-body problems more generally, the baseline approach for collective neutrino flavor oscillations 
is a mean-field model, replacing two-body interactions with an average one-body potential~\cite{Pantaleone92,Pantaleone:1992eq,Sigl:1993fr,Qian95,McKellar:1994uq,Balantekin:2006tg}.
To study the beyond-mean-field time evolution of interacting neutrinos, the system (in the two-flavor, single angle approximation) was mapped~\cite{Patwardhan:2019zta,Cervia:2019res} to the Richardson-Gaudin Hamiltonian, which was originally 
developed as a model of pairing solvable by the Bethe \textit{ansatz} and which has since been applied to a variety 
of many-body systems, including atomic nuclei~\cite{RevModPhys.76.643}.

For neutrino numbers $N \geq 10$, however, numerical solutions of the Bethe \text{ansatz} for time evolution of the 
many-neutrino system were empirically unstable.
As a result, the authors of Ref.~\cite{Patwardhan:2021rej} instead utilized a fourth-order Runge-Kutta (RK4) method to integrate the 
time-dependent Schr\"odinger equation for the $N$-body neutrino wave function. By using sparse-matrix representations of the operators constituting the Hamiltonian, neutrino systems with $N$ up to $16$ could be studied. The exponential scaling of the computational complexity eventually renders this method intractable for larger numbers of neutrinos. On the other hand, this work made a potentially useful observation, namely that the degree of quantum entanglement seemed to be larger among the neutrinos nearer to the \lq\lq spectral splits\rq\rq\ in their energy distributions, and smaller among the neutrinos away from these splits. This suggests that a numerical scheme that can zero in on specific subregions of the full Hilbert space where the entanglement primarily resides could potentially yield accurate results while scaling more favorably with $N$, compared to traditional integration methods.

Given the limitations of both the Bethe \textit{ansatz} and direct RK4 approaches, 
in this paper we turn to the use of tensor network methods to model correlated neutrino wave functions and to investigate the dynamics of collective neutrino oscillations.  Tensor networks provide a means for calculating dynamics using a truncated basis set with dimensions that can grow subexponentially with system size but that can nonetheless  be highly entangled. 
In this approach, the many-body wave function is written in terms of inner products of tensors that encode pairwise entanglement.

The problem of collective neutrino oscillations involves {nonlocal} (in momentum space) interactions, and so it is nontrivial to determine whether the method is well-behaved in such a way that one may practically apply these recent methods without requiring exponential growth in the \lq bond dimension\rq\ {[as defined in Eq.~\eqref{eq:left-canon}]} used to forward-integrate the wave function.
In order to treat the nonlocal Hamiltonian, one can implement the tensor network using SWAP operations 
(analogous to SWAP gates in quantum computing, both defined in Ref.~\cite{Roggero:2021asb}) to ``localize'' the interactions~\cite{Roggero:2021asb,Roggero:2021fyo,Martin:2021bri}: a nonlocal interaction is replaced by 
interactions between network neighbors interlaced with the SWAP operations that make distant network neighbors appear 
nearby.
Tests of this method have been limited, however, to just a few different neutrino {momenta}, typically while employing the two-beam model~\cite{PhysRevD.88.045031}. 

While intriguing properties such as phase transitions  can still be learned in a model with a reduced set of neutrino momenta, including many neutrino modes leads to additional effects, such as spectral splits, which we expect to be 
physically relevant to astrophysical phenomena {in these environments}. To address these issues,
in this paper we calculate the time evolution of a tensor network model of the many-neutrino wave function using 
 recent time-dependent variational principle (TDVP) methods~\cite{Haegeman_2016,PhysRevB.102.094315}.  We compare 
 our reduced-basis, tensor-network model 
 against two methods calculating the entire wave function: fourth-order Runge-Kutta and Lanczos propagation. 
 (Lanczos  is a different kind of reduced-basis method: while the underlying basis dimension is not 
 changed, Lanczos efficiently computes time evolution by projecting into a small effective basis dictated by 
 the initial state and by a few iterated applications of the Hamiltonian.)
 Furthermore, since these numerically exact methods use the entire basis and thus scale exponentially  with  the system size $N$, there is a limit in size after which the accuracy of tensor network methods cannot be assessed by comparison with other methods; as such, we introduce a strategy for consistency checks in our evolved wave function to guide our calculations for $N$ beyond the abilities of RK and Lanczos methods on modern hardware.

\begin{figure}[htbp]
    \includegraphics[width=0.99\linewidth]{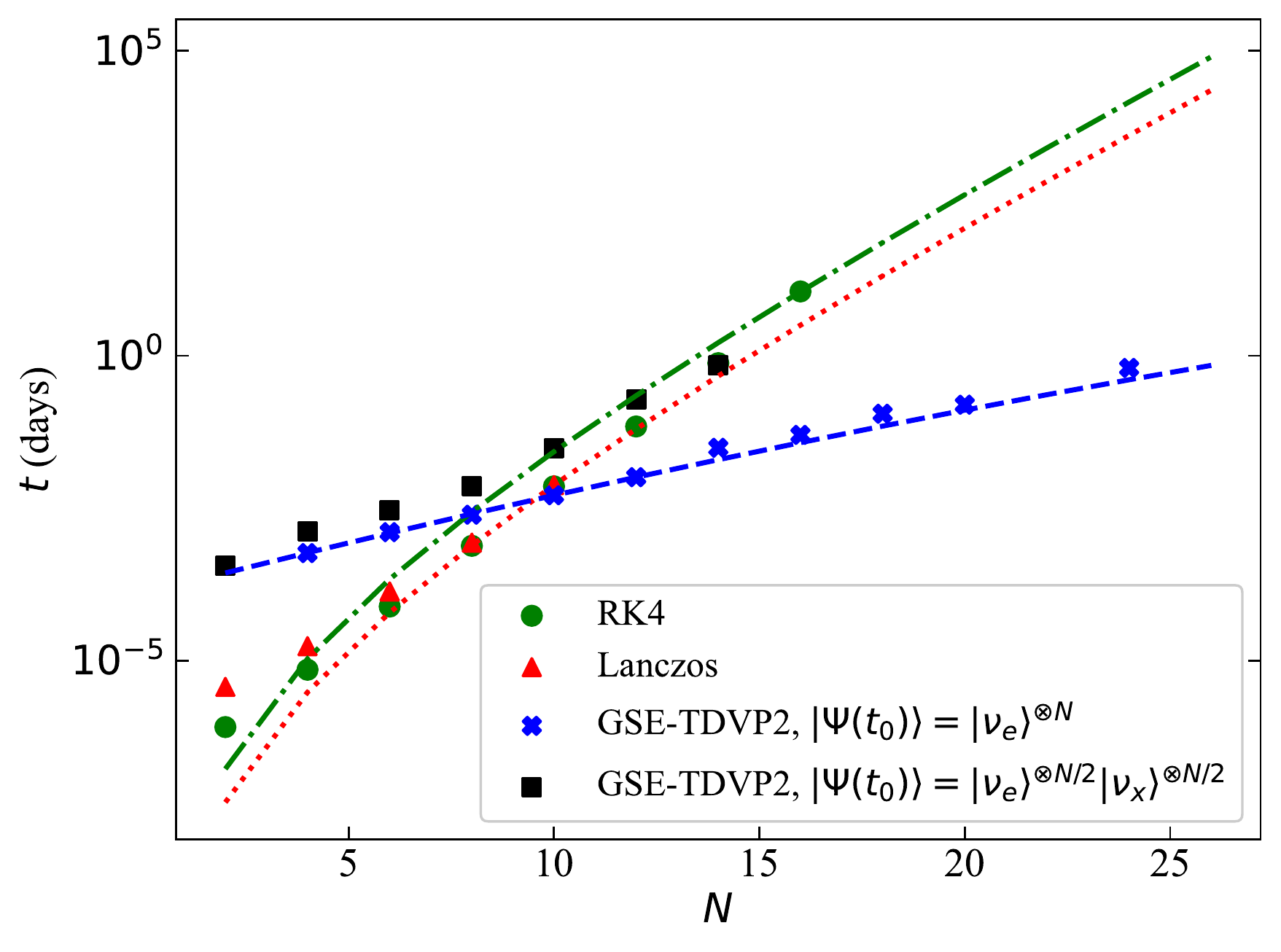}
    \caption{Computation times of different numerical simulations (see Sec.~\ref{sec:methods} for details) of the flavor evolution of the $N$-neutrino system.  
    Tensor networks evolved under a time-dependent variational principle (i.e., GSE-TDVP2) permit substantial speedup in larger-$N$ calculations for collective oscillations albeit depending on initial conditions, 
    while computational time for RK4 and Lanczos is nearly independent of the initial state.
    In the case of GSE-TDVP2, we specify the initial state as $\ket{\nu_e}^{\otimes N}$ and $\ket{\nu_e}^{\otimes N/2}\ket{\nu_x}^{\otimes N/2}$, respectively. 
  Calculations  performed on a single CPU ($2.5$ GHz {Quad-Core Intel Core i7} processor) for comparison; the {\small{ITensor}} library supports OpenMP multithreading, allowing reduced wall-clock  time.
    For tensor network calculations used  in this plot, we require the use of the lowest bond dimensions and largest time steps that permit the error in each calculation, as quantified in Sec.~\ref{sec:ehror}, to be as small as that in the RK4 and Lanczos propagation calculations for the same initial condition. Fit functions for all methods  described in Sec.~\ref{sec:results}.}
    \label{fig:time}
\end{figure}

We find that tensor network calculations are potentially very useful for the study of the collective neutrino oscillation problem. In particular, for initial conditions of a spectrum of neutrinos that result in fewer different spectral split frequencies, the entanglement of the system can be efficiently described by a matrix-product state (MPS) representation of our many-body neutrino state, permitting {memory-} and time-efficient computations of time evolution. As depicted in Fig.~\ref{fig:time}, we find that for an initial condition with just one spectral split (i.e., a system starting with all electron-flavor neutrinos), the basis set for representing our MPS wave function can be reduced greatly, allowing for improvements in complexity. In contrast, in the case of a system with two spectral splits,  though for sufficiently large $N$ we find some speedup, the required basis set to obtain accuracy comparable to exact methods is much greater in size. We conclude the usefulness of this treatment of neutrinos depends considerably on the number of spectral splits that result from an initial condition. {Nonetheless, computing the time evolution of the many-neutrino systems using a time-dependent variational 
principle on a tensor network---the central approach of this paper---is a promising tool for modeling and 
understanding the beyond-mean-field behavior of collective flavor oscillations. }

\subsection{Outline of the paper}

The paper is organized as follows. In Sec.~\ref{sec:def}, we introduce our toy model of collective neutrino oscillations in a dense neutrino gas. In Sec.~\ref{sec:methods}, we briefly summarize the methods for calculating the entire wave function after time evolution from the time-dependent Hamiltonian describing our problem, and we introduce the recent tensor network techniques for treating the same problem. In Sec.~\ref{sec:ehror}, we define a measure for error in our evolved wave function calculated with any method, based upon the instantaneously conserved charges of the Hamiltonian from Ref.~\cite{Pehlivan:2011hp}, to assess how precisely the state solves the Schr\"odinger equation. In Sec.~\ref{sec:results}, we use this new measure to assess the quality of a given calculation and decide upon appropriate time-step sizes as well as bond dimension values set, depending upon the initial condition chosen for our problem. In Sec.~\ref{sec:conclusion}, we summarize our findings and suggest paths forward to investigate larger systems within our model. Finally, in Appendix~\ref{sec:exact-details} we provide additional details about the calculations of entire wave functions, while in Appendix~\ref{sec:tensor_net_form}, we explain in greater detail the tensor network methods to perform time evolution with long-range interactions.

\section{Definitions}
\label{sec:def}

To study collective neutrino oscillations from a many-body perspective in a two-flavor, single-angle model, we consider the Hamiltonian~\cite{Balantekin:2006tg,Pehlivan:2011hp,Birol:2018qhx,Balantekin:2018mpq,Cervia:2019nzy,Patwardhan:2019zta,Cervia:2019res,Patwardhan:2021rej}
\begin{equation}
    H(t) = -\sum_{\omega}\omega J^z_\omega + \mu(t) \sum_{\substack{\omega,\omega'\\\omega'\neq\omega}} \vec{J}_\omega\cdot\vec{J}_{\omega'},
    \label{eq:saham}
\end{equation}
where $\omega$ denotes the vacuum oscillation frequencies of the neutrinos and $\mu(t)$ is the time-dependent, {angle-averaged} $\nu$-$\nu$ interaction strength. Here, we have used the $SU(2)$ neutrino isospin operators (in the mass basis) $\vec{J}_\omega$ for a neutrino of a given mode $\omega$:
\begin{align}
    J_\omega^z &= \frac12 (c_{1\omega}^\dagger c_{1\omega} - c_{2\omega}^\dagger c_{2\omega}),\\
    J_\omega^+ &= c_{1\omega}^\dagger c_{2\omega} = (J_\omega^-)^\dagger,
\end{align}
with $c_{i\omega}^\dagger$ and $c_{i\omega}$ as the (fermionic) creation and annihilation operators of a neutrino of mass eigenstate $\ket{\nu_i}$ in the mode {$\omega =\Delta m^2/(2|\mathbf{p}|)$, where $\Delta m^2$ is the mass-squared difference and $\mathbf{p}$ is the momentum of this mode. Each neutrino in this model therefore has a description as a plane wave with well-defined momentum. Such a treatment is considered adequate for capturing the coherent many-body effects in collective neutrino oscillations~\cite{Friedland:2003eh,Friedland:2006ke}. }

{In the single-angle approximation (e.g., Ref.~\cite{Duan06a}), all the time dependence of the Hamiltonian is described by a single, angle-averaged parameter $\mu(t)$. Consequently, the flavor evolution of a neutrino in this approximation depends only on its energy and not on the direction of its momentum, reducing the computational complexity of the problem. The single-angle approximation is known to exhibit many of the same collective phenomena, such as synchronized precession and spectral swaps/splits, that are also present in the more sophisticated, multiangle treatments~\cite{Duan06b}.}


The many-body state $\ket{\Psi(t)}$ of a $N$-neutrino system then evolves according to the Schr\"odinger equation\footnote{Note that there is just one affine parameter in the evolution equation, if the neutrinos are assumed to be relativistic and their emission from the source is assumed to be time-independent. The latter assumption can be justified based on an observed hierarchy in the dynamical timescales. In a core-collapse supernova environment, a neutrino would experience significant interactions with other neutrinos over an interval of $\lesssim 1000$\,km~\cite{Duan:2010fr}, or equivalently, a timescale of $\mathcal O(\mathrm{ms})$. This is much smaller than the $\mathcal O(\mathrm{s})$ timescales over which the emission characteristics (luminosities and energy spectra of different flavors) change significantly in the late-time neutrino-driven wind phase; see, e.g., the estimates in Refs.~\cite{Burrows:1984zz,Burrows:1990ts,Janka:2017vlw} for the cooling and deleptonization timescales. Throughout the text, the affine parameter is henceforth interchangeably referred to as either time $(t)$ or radius $(r)$.}~with the time-dependent Hamiltonian in Eq.~\eqref{eq:saham}:
\begin{equation}
    i\frac{\mathrm{d}}{\mathrm{d}t}\ket{\Psi(t)} = H(t)\ket{\Psi(t)}.
    \label{eq:schrod}
\end{equation}

The polarization vector for a neutrino with a given $\omega$ is defined as
\begin{equation}
    \vec P(\omega) = 2 \bra{\Psi} \vec J_{\omega} \ket{\Psi},
    \label{eq:mb-polar}
\end{equation}
where $\vec{J}_{\omega}$ is the corresponding isospin operator for that neutrino and $\Psi$ is the many-body wave function of the $N$-neutrino system. 
The polarization vectors can also be obtained from the Pauli spin decomposition of the individual neutrino density matrices in the mean-field limit, or the \lq\lq reduced\rq\rq\ density matrices in the case of a many-body calculation. In each case, the decomposition is given by
\begin{equation}
    \rho(\omega) = \frac12 ( \mathbb{1} + \vec{\sigma} \cdot \vec{P}(\omega) ),
\end{equation}
where $\mathbb{1}$ is the $2 \times 2$ identity matrix and $\sigma_j$ are the Pauli spin matrices. From the above expressions, one can also conclude that the probability of finding an individual neutrino in the mass eigenstate $\ket{\nu_1}$ is
\begin{equation}
    P_{\nu_1}(\omega) = \frac12(1+P_z(\omega)) = [\rho(\omega)]_{11},
    \label{eq:polarprob}
\end{equation} 
i.e., the $11$ matrix element of the (reduced) density matrix, $\rho(\omega) = \mathrm{Tr}_{\omega'(\neq\omega)}[\ket{\Psi}\!\bra{\Psi}]$.

Then, entanglement entropy between a neutrino with frequency $\omega$ and the rest of the ensemble can be obtained from 
\begin{align}
    S(\omega) &= -\sum_{s=\pm}\lambda_s(\omega)\log[\lambda_s(\omega)],
    \label{eq:entropy}
\end{align}
where
\begin{align}
    \lambda_\pm(\omega) &= \frac{1}{2}(1\pm|\vec{P}(\omega)|)
    \label{eq:1dens-eigen}
\end{align}
are the eigenvalues of the reduced density matrix $\rho(\omega)$. 
Reference~\cite{Patwardhan:2021rej} highlights the relationship between entanglement in an evolved neutrino many-body system and the spectral split features  of the neutrino spectrum. The spectral split is a feature of interacting neutrino systems whereby the neutrino survival and conversion probabilities exhibit a transition about a split frequency $\omega_s$~\cite{Duan06a,Duan06b,Raffelt07,Raffelt:2007,Duan07a,Duan07b,Duan07c,Fogli07,Duan:2008qy,Dasgupta:2008qy,Dasgupta:2009mg,Dasgupta:2010cd,Friedland:2010yq,Galais:2011gh,Pehlivan:2016lxx,Birol:2018qhx}. Though this phenomenon is predicted also in the mean-field limit of this system, a many-body description reveals that $S(\omega)$ is greatest for $\omega\sim\omega_s$~\cite{Patwardhan:2021rej}.

\section{Methods}
\label{sec:methods}


As per the bulb model~\cite{Duan:2010fr}, we take a system composed of neutrinos in definite flavor states emitted isotropically from a source; in this case, the initial many-body state has the form $\ket{\Psi}=\bigotimes_{j=1}^N\ket{\nu_{\alpha_j}}$, where $\alpha_j=e$ or $x$ for each $i$, and evolves according to Eq.~\eqref{eq:schrod} with the time-dependent Hamiltonian in Eq.~\eqref{eq:saham}. Additionally, we consider a ``box spectrum'' with discrete, equally spaced vacuum oscillation frequencies $\omega_j = j\omega_0$, for $j=1,\ldots,N$ (where $\omega_0$ is an arbitrary reference frequency), such that each oscillation frequency is occupied by a single neutrino. 
{Similarly in keeping with Ref.~\cite{Cervia:2019res}, we use the mixing angle $\theta=0.161$, a single-angle coupling 
\begin{equation}
    \mu(r) = \mu(R_\nu)\bigg[1-\sqrt{1-\bigg(\frac{R_\nu}{r}\bigg)^2}\bigg]^2
\end{equation}
with $R_\nu = 32.2\:\omega_0^{-1}$ and $\mu(R_\nu)= 3.62\times10^4\:\omega_0$, and an initial time/radius given by $\mu(r)=5\:\omega_0$.}

\subsection{Mean-field theory}
\label{sec:mft}

Within mean field theory, the wave function is always considered a direct product of individual neutrino wave functions, i.e., $\ket{\Psi} = \bigotimes_\omega \ket{\psi(\omega)}$, and Eq.~\eqref{eq:mb-polar} reduces to $\vec P(\omega) = 2 \bra{\psi(\omega)} \vec J_{\omega} \ket{\psi(\omega)}$ (e.g., Ref.~\cite{Balantekin:2006tg}). 
Equivalently, we may write the density matrix of each neutrino as simply $\rho(\omega) = \ket{\psi(\omega)}\!\bra{\psi(\omega)}$ without explicitly performing a trace over the local Hilbert spaces of all other neutrinos $\omega'\neq\omega$. 
As a consequence, in the mean-field case, $|\vec{P}(\omega)|=1$ for each neutrino, implying $S(\omega)=0$ exactly. In contrast, when the neutrino mode $\omega$ is maximally entangled with its environment (which in this case, is composed of all the other neutrinos), $|\vec{P}(\omega)|=0$, and so entanglement entropy $S(\omega)=\log(2)$. In this sense, entanglement entropy serves as a probe of many-body deviations from the mean-field theory. 

Moreover, in the mean-field treatment, the evolution of the $N$-body neutrino system can then be described using a set of $N$ differential equations, each describing the evolution of one neutrino. In terms of the polarization vectors $\vec{P}(\omega)$,\footnote{Each polarization vector has dependence on $r$, suppressed in our notation above for brevity.} the evolution equations can be written as
\begin{equation}
    \frac{\mathrm{d}\vec P(\omega)}{\mathrm{d}t} = \omega \vec B \times \vec P(\omega) + \mu(r) \left[ \sum_{\omega'} \vec P(\omega') \right] \times \vec P(\omega),
\end{equation}
where in the mass basis $\vec B = (0,0,-1)$.

\subsection{Numerical calculations of time evolution of many-body wave functions}
\label{sec:methods-exact}

{One can of course directly solve the time-dependent Schr\"odinger  Eq.~\eqref{eq:schrod}  in the Hilbert space for $N$ neutrinos spanned by $2^N$ basis states of 
the form $\bigotimes_{\omega}\ket{\nu_{\alpha_\omega}}$ , where each $ \ket{\nu_{\alpha_\omega}} $ 
is a flavor-spinor for neutrino frequency $\omega$ {(e.g., in the flavor basis, each $ \nu_{\alpha_\omega} = \nu_e$ or $\nu_x$, resulting in $2^N$ combinations over $N$ frequencies)}.
While   limited by the exponential growth of the basis dimension, this is nonetheless 
a useful benchmark for other methods.
While large in dimension, because the Hamiltonian from Eq.~\eqref{eq:saham} has interactions 
at most between two flavor-spinors, the matrix representation of $\hat{H}$ in the $e,x$ basis 
(equivalent to $\uparrow,\downarrow$ for ordinary spinors) is sparse.
}

In Ref.~\cite{Patwardhan:2021rej}, we 
evolve the many-body state $\ket{\Psi(t)}$ via the classical RK4, {a textbook~\cite{press1996numerical} approach to solving ordinary differential equations.}
{The goal of this effort was} to extend  to $N\geq10$ earlier calculations~\cite{Cervia:2019res} {that were performed by diagonalizing $H$ efficiently via Bethe ansatz while studying the behavior of instantaneously conserved quantities of the system}. 

Another computational approach to numerically time-evolving the many-body state $\ket{\Psi(t)}$, working still in a sparse-matrix representation, is a Lanczos propagation~\cite{park1986unitary,hochbruck1997krylov,mohankumar2006time} for a time-dependent Hamiltonian~\cite{cremon2013adaptive}. 
The Lanczos algorithm~\cite{stewart2001matrix},  a dimensional reduction method which generates the basis vectors of an effective (Krylov) 
subspace by repeated application of the Hamiltonian,  is widely used, particularly in nuclear structure physics~\cite{RevModPhys.77.427}.
In our application, the many-body state is forward-integrated according to Eq.~\eqref{eq:schrod} by applying a time-evolution operator;
\begin{align}
    \ket{\Psi(t+\delta t)} = U(t+\delta t;t)\ket{\Psi(t)}. 
    \label{eq:time-evo-op-step}
\end{align}
{While formally one should compute the time-evolution operator $U$ 
in a Magnus expansion 
(see Appendix~\ref{sec:exact-details} for details), 
in practice we found  the naive time-evolution operator $\exp(-i H(t) \delta t)$, was sufficient.  The Lanczos algorithm aids the efficient calculation of $U$ by 
exponentiating the Hamiltonian projected into a very small but effective subspace, 
generated by  applying powers of $H$ on the state $| \Psi(t)\rangle$---again, details can 
be found in Appendix~\ref{sec:exact-details}.} 


Using established methods presented in Ref.~\cite{Patwardhan:2021rej} for evolving a many-body state with the Hamiltonian $H(t)$, one can verify for $N=2$ to $16$ that using Lanczos propagation 
to approximate the evolved state to order $(\delta t)^5$ with the appropriately chosen $\delta t$ produces accurate results for the wave function even after evolving over many time steps. When compared with results from RK4, the value of each coefficient in the wave function, $\braket{j|\Psi(t)}$ ($j=0,\ldots,2^N-1$), was calculated with a discrepancy $\lesssim10^{-3}$. 
{Because these methods are in numerical agreement with one another, we do not separately show explicit results for the time evolution of the wave function in the case of the Lanczos   method.} 
Note that the truncations involved in these numerical methods can cause the normalization of the resulting wave function to change, so one may need to normalize the resulting wave function between time steps. 
With each of these methods, we prescribe the time step to scale in $N$ as inversely with the scaling of the difference between the extremal eigenvalues in our Hamiltonian: $\delta t\sim 0.1[\mu \frac{N}{2}(\frac{N}{2}+1)+\sum_\omega|\omega|]^{-1}$, where $\mu$ is evaluated at the radius prior to taking this time step. 
Just as with the use of RK4, we implement Lanczos propagation in a sparse-matrix representation, permitting calculations of the evolved many-body wave function according to a time-dependent Hamiltonian for up to $N=16$ on a personal computer. In implementing the sparse representation in our own programs, we make use of submodules from {\small{SPARSKIT}}~\cite{Saad94sparskit:a}, a Fortran90 library for performing operations with sparse matrices. 

However, because the number of nonzero elements in the many-body Hamiltonian matrix grows as $O(N^2 2^N)$, memory limitations severely restrict the values of $N$ that can be studied. The time required to calculate the time evolution using these methods also grows exponentially in $N$.  Tensor networks provide a method that in principle could scale more favorably with $N$.  We describe these methods in the next subsection and investigate how the resources needed to obtain accurate results using tensor networks scale with $N$ in the next section. 

\subsection{Calculating matrix product state wave functions}
\label{sec:methods-mps}

This exponential growth in problem size motivates the use of tensor network methods; appropriately chosen tensor network representations allow for the complexity of the problem to scale instead much more slowly with system size. However, it is not clear \emph{a priori} how the size of the tensor network representation needed to obtain accurate results scales with $N$.  Determining whether the necessary size grows slower than exponentially with system size is a key goal of this paper. The bond dimension required to obtain accurate results may scale either exponentially or polynomially for our system; using the methods that follow, we will in particular investigate how methods based upon the time-dependent variational principle   may scale with $N$ in treating our problem.

Here, we briefly outline how to express the many-body wave function and operators acting upon this state in terms of MPSs, followed by a sketch of the TDVP algorithm we use, reserving greater detail for Appendix~\ref{sec:tensor_net_form}. 
The mathematical language associated with this decomposition will be referred to as ``matrix product'' or ``tensor network'' formalism interchangeably for our purposes. After establishing the mathematical definitions for describing a MPS, we will briefly outline the computational process of time evolving a wave function efficiently in a closed quantum system in the matrix product formalism, using the TDVP in the tensor network formalism. 
To this end, we largely adopt the language used in Ref.~\cite{Haegeman_2016}, which first outlined the version of a TDVP algorithm that we use in this work, to describe MPS formalism and its use with the TDVP. 
For a more general review of time evolution methods utilizing MPS representations of wave functions, see, e.g., Ref.~\cite{PAECKEL2019167998}.

\subsubsection{General MPS review}
\label{sec:methods-mps-formal}

Let us begin by establishing the language needed to describe a MPS. 
For a system of $N$ neutrinos where we bin the spectrum of $\omega$ such that all neutrinos have distinct frequencies, we may label their frequencies with index values $1,\ldots,N$; in the context of tensor network formalism, we can refer to these frequencies interchangeably as ``sites.'' (Put another way, while the MPS community frequently considers sites in reference physical locations along a lattice, we are instead considering sites in reference to definite momentum states for different neutrinos in a spectrum.) 
Then, given a wave function $\ket{\Psi}$ decomposed in the flavor basis
\begin{equation}
    \ket{\Psi} = \sum_{\alpha_1,\ldots,\alpha_N=e,x} \Psi^{\alpha_1\cdots\alpha_N}\ket{\nu_{\alpha_1}\cdots\nu_{\alpha_N}},
\end{equation}
we may view $\Psi^{\alpha_1,\ldots,\alpha_N}=\braket{\nu_{\alpha_1},\ldots,\nu_{\alpha_N}|\Psi}$ as a complex-valued tensor with $N$ indices each spanning a two-dimensional vector space. By $N-1$ iterations of Schmidt decomposition (see, e.g., Refs.~\cite{Nielsen:2011:QCQ:1972505,2011AnPhy.326...96S,Roggero:2021asb}) 
starting from the leftmost indices, we may write this component as a product of site-dependent matrices $\psi_L^{\alpha_j}(j)$:
\begin{align}
    \Psi^{\alpha_1\cdots\alpha_N} &=
    \psi_L^{\alpha_1}(1)\cdots \psi_L^{\alpha_N}(N)
    \nonumber \\
    &=\sum_{\beta_1=1}^{D_1}\cdots\sum_{\beta_{N-1}=1}^{D_{N-1}} \psi_{L,\beta_1}^{\alpha_1}(1) \psi_{L,\beta_1\beta_2}^{\alpha_2}(2) \cdots
    \nonumber \\ &\phantom{=\sum_{\beta_1=1}^{\chi_1}}\times\psi_{L,\beta_{N-2}\beta_{N-1}}^{\alpha_{N-1}}(N-1) \psi_{L,\beta_{N-1}}^{\alpha_N}(N),   
    \label{eq:left-canon}
\end{align}
where for fixed $(\alpha_1,\ldots,\alpha_N)$: $\psi_L^{\alpha_1}(1)$ is a dim-$D_1$ row vector, $\psi_L^{\alpha_j}(j)$ for $1<j<N$ are $D_{j-1}\times D_{j}$ rectangular matrices, and $\psi_L^{\alpha_N}(N)$ is a dim-$D_{N-1}$ column vector. This general decomposition into a matrix product is also referred to as an example of a ``tensor train,'' and specifically with our choice of direction in Schmidt decomposition is a ``left-canonical'' form for the tensor.  Here, we call 
$\beta_j$ and 
$D_j$ the 
``bond indices'' and 
``bond dimensions'' of our tensor train. 
An exact representation of $\Psi$ is obtained if we take 
$D_j=\min\{2^{j},2^{N-j}\}$. 

With these exact choices for maximal bond dimensions, our procedure 
requires computational resources that scale exponentially with $N$. To reduce the scaling of computational resources with $N$, we seek to allow the maximum values of $D_j$ used in our computations to grow minimally with $N$ while maintaining a similar level of error as obtained in methods such as RK4. 
Note that the bond dimension can help us to assess the entanglement in our ensemble; in the case that there is exactly zero entanglement entropy at each site, we find that $D_j=1$ for each $j$ permits an exact representation of the state, resulting in an independent dim-2 vector subspace for each body---just as we would write in the mean-field theory calculations of the ensemble state [i.e., $\psi^{\alpha_j}(j)=\braket{\alpha_j|\psi(\omega_j)}$ in this case]. Regardless of our particular choice of $D_j$, we order the singular values of each tensor by size and keep the $\leq D$ largest values.

\subsubsection{TDVP for a neutrino MPS}
\label{sec:methods-mps-tdvp}

Numerous recent developments have been made in the community studying time evolution of MPS representations of spin systems, whose accuracy is controlled in part by the choice of maximum bond dimension $D$ use for a given calculation. 
In particular, in connection with MPS density matrix renormalization group (DMRG) techniques, Refs.~\cite{Haegeman:2011zz,Haegeman_2016,PhysRevB.101.235123} developed a method of real-time evolution in analogy with the TDVP. This TDVP algorithm in particular readily permits calculations with a spin Hamiltonian involving nonlocal interactions, with an acceptable level of accuracy reproduced for the case of a power-law potential~\cite{Haegeman_2016,PhysRevB.101.235123}. 
Viewing $\psi_{\beta_1}^{\alpha_1}(1),\ldots,\psi_{\beta_{N-1}}^{\alpha_N}(N)$ from Eq.~\eqref{eq:left-canon} as $N$ coordinates parametrizing a MPS manifold $M$ of the state $\ket{\Psi}$, the TDVP can be interpreted geometrically as a projection of the right-hand side of the Eq.~\eqref{eq:schrod} onto the tangent space of said manifold at a location given by $\Psi$, $T_\Psi M$, resulting in the nonlinear differential equations
\begin{equation}
    i\diff{}{t}\ket{\Psi(t)} = P_{T_\Psi M}H(t)\ket{\Psi(t)},
\end{equation}
where $P_{T_\Psi M}$ is the projection operator onto the tangent space. Specifically, we may choose different projection operators such that we evolve only one or multiple tensors $\psi(j)$ in Eq.~\eqref{eq:left-canon} at once; their forms and the consequences of each choice are also presented in Appendix~\ref{sec:tensor_net_form}. The TDVP algorithm implemented with the choice of having $n$ active sites being evolved at once is referred to as $n$TDVP, with $n=1,2$ generally found to be practical computationally. 

Furthermore, an augmentation called global subspace expansion (GSE) has more recently been made to the $n$TDVP algorithm~\cite{PhysRevB.102.094315}, the combination of which we call GSE-TDVP$n$, where $n$ is the number of active sites being evolved in each TDVP step. In this procedure, one includes additional singular values from global Krylov vectors, calculated as $[1-i\delta tH(t)]^\ell\ket{\Psi(t)}$ ($\ell\in\mathbb{N}$), into the bonds of the MPS wave function obtained between time steps of TDVP (in analogy with DMRG to optimize for a mixture of lowest-lying energy eigenstates). It was found that this addition provided greater flexibility to the choice of appropriate time-step sizes used for time evolution in cases such as the one-axis twisting model. However, the problem of collective neutrino oscillations in principle requires not only nonlocal interactions, 
but also the inclusion of one-body kinetic terms in the plane-wave treatment of neutrinos in our toy model. As such, it is not immediately apparent that entanglement describing the evolved many-body state of our system is accurately captured by these recent methods without requiring exponential growth in the bond dimension used to forward-integrate the wave function.


Notably, each virtual bond within the tensor train need not have identical dimension $D_j$ ($j=1,\ldots,N-1$). In the repeated singular value decomposition to obtain a MPS (e.g., outlined in Refs.~\cite{2011AnPhy.326...96S,Roggero:2021asb}), the bonds closest to the ends of the $\omega$ spectrum would have dim $\leq 2$ while bonds closest to the center have dimension $\lesssim 2^{N/2}$. In 1/2TDVP algorithms made available through the {\small{TeNPy}} library~\cite{tenpy} as well as GSE-TDVP1/2 algorithms in the {\small{ITensor}} library~\cite{itensor,PhysRevB.102.094315}, we can control the maximum cutoff dimension for all of the bonds, which we denote by $D$. 
However, with an initial wave function in which neutrinos are unentangled, carefully note that 1TDVP calculations  prevent the bond dimensions in the initial MPS from rising at all as time evolves, and correspondingly entanglement entropy is negligible throughout the calculation. Consequently, a one-site effective Hamiltonian calculation with an entirely fixed bond dimension can only replicate the results of an exact many-body calculation by beginning a time evolution using 2TDVP or GSE-TDVP1/2 for long enough to let all bonds reach their maximum permitted dimensions before then switching to 1TDVP. 
Moreover, we find that the greatest flexibility in the choice of time-step size and bond dimension, which are determined by a procedure described in Sec.~\ref{sec:results}, is afforded by GSE-TDVP2. Therefore, this algorithm will be the MPS time evolution method used throughout our results in that section. 

While Ref.~\cite{Roggero:2021asb} finds that a bond dimension $D$ that scales linearly with the number of neutrinos is adequate in studying a two-beam model of collective oscillations, we find that the scaling of $D$ with $N$ is more complex for our case.
By calculating the magnitude of the deviations of the results of our calculations from satisfying Ehrenfest's theorem, as outlined in Sec.~\ref{sec:ehror}, we can evaluate how much this restriction of bond dimension impacts the precision of the TDVP evolution of the wave function. 

Beside choosing a bond dimension cutoff $D$ in our tensor network calculations, we must also take care in choosing a time-step size $\delta t$ throughout the evolution of our many-body wave function. As we will demonstrate in our Results, the problem of determining an appropriate $\delta t$ to accurately evolve our MPS wave function is not entirely straightforward; while smaller time steps may help to more accurately forward-integrate our evolution equations, there are errors from ignoring singular values in both deriving our equations and following each time step, implying that shrinking $\delta t$ to be too small can introduce even greater errors{, as has been described, e.g., in Ref.~\cite{PAECKEL2019167998}}.\footnote{As a brief remark, it is worth pointing out that this growth in error with number of time steps, or, conversely, with decreasing 
step size, is not unique: it is well known, 
for example, in numerical Lax methods~\cite{press1996numerical}.} (For a more detailed explanation, see Appendix~\ref{sec:tensor_net_form}.) In general, one must select a way to assess the error of a method without already knowing the exact solution to the problem; we shall propose a method for our problem in the following section. 
However, for initial conditions where entanglement is limited (i.e., $\ket{\nu_e}^{\otimes N}$), one {can} determine an appropriate $\delta t$ in tensor network calculations by evolving $\ket{\Psi(t_0)}$ via 2TDVP to $t=t_0+\omega_0^{-1}$ using decreasing step sizes. Comparing the evolved wave function $\ket{\Psi_n}\equiv\ket{\Psi(t_0+2^n\delta t_n)}$ obtained with each step size $\delta t_n\equiv2^{-n}\omega_0^{-1}$, we choose $\delta t=\delta t_n$ for the smallest natural number $n$ such that $1-|\!\braket{\Psi_{n+1} | \Psi_{n}}\!|^2$ is less than some chosen tolerance value. Comparison with results obtained using sparse matrix computations in the complete Hilbert space for $N\leq16$ suggests that a tolerance of $10^{-4}$ is appropriate for finding a practical $\delta t$ in TDVP calculations of our system. For more general initial conditions and when using GSE in addition to TDVP, we will see below that determining an appropriate choice of $\delta t$ will require greater attention. {In fact, by carefully checking how error accumulates with differing $\delta t$, we find that certain larger values are often preferred in accurately evolving the MPS wave function than the upper bound prescribed for RK4 or Lanczos propagation by the argument outlined in Sec.~\ref{sec:methods-exact}. }

\section{Consistency checks for numerical treatments}
\label{sec:ehror}

To compare tensor network methods to the other methods described above, we need to consider both how the resources needed to calculate the evolution scale with system size and whether the accuracy of the solution that is obtained is adequate.
Because the memory used and the computation time required grow polynomially with $D$~\cite{Haegeman_2016} yet the maximum physical choice of $D$ can grow exponentially in $N$, it is important to characterize the accuracy of the calculation when $D$ is modest. 
For small $N$ the results of the tensor network method can be compared to the numerically exact results obtained by the other methods, but for larger $N$ it is useful to have another method to assess the accuracy of the results.
In this section we discuss consistency checks that follow from conservation laws that can be used to assess whether the results yielded by the tensor network method with a given bond dimension are accurate.

It is known that the many-body neutrino Hamiltonian in the single-angle approximation has a number of commuting invariant operators. One such operator is $J^z = \sum_\omega J^z_\omega$, i.e., the $z$-component of the total neutrino isospin in the mass basis. Another set of invariants, given by~\cite{Pehlivan:2011hp} and used further by \cite{Birol:2018qhx,Cervia:2019nzy,Patwardhan:2019zta} is
\begin{equation}
    h_\omega = -J^z_\omega + 2\mu \sum_{\omega' (\neq \omega)} \frac{\vec{J}_\omega \cdot \vec{J}_{\omega'}}{\omega - \omega'}.
    \label{eq:homega}
\end{equation}
These invariants can be used as consistency checks in numerical calculations. We do so by using Ehrenfest's theorem, which states that the time evolution of the expectation value of an operator $A$ is given by\footnote{Using the chain rule of differentiation, one can write
\begin{equation}
    \diff{\langle A \rangle}{t} = \left\langle \diffp{A}{t} \right\rangle + \left(\diff{}{t}\bra{\Psi}\right)A\ket{\Psi} + \bra{\Psi} A \left(\diff{}{t}\ket{\Psi}\right).
\end{equation}
For a wave function $\ket\Psi$ which satisfies the Schr\"odinger equation with a Hamiltonian $H$, one can then use the Hermiticity of $H$ to obtain Eq.~\eqref{eq:ehrenfest}.
}
\begin{equation}
    \diff{\braket{A}}{t} = \frac{1}{i} \braket{[A, H]} + \left\langle \diffp{A}{t} \right\rangle,
    \label{eq:ehrenfest}
\end{equation}
where the expectation values $\langle \, \cdot \, \rangle$ are calculated with respect to a wave function that satisfies Eq.~\eqref{eq:schrod}. 
In particular, when $A$ is an invariant of the Hamiltonian, i.e., if $[A,H] = 0$, one has
\begin{equation} 
     \diff{\braket{A}}{t} - \left\langle \diffp{A}{t} \right\rangle = 0,
    \label{eq:ehrenfestzero}
\end{equation}
for a wave function $\ket\Psi$ that satisfies the Schr\"odinger equation. As an example, taking $A = J^z$ in the above equation gives the simple result $\mathrm{d}\langle J^z \rangle/\mathrm{d}t = 0$, since $J^z$ has no explicit time dependence. Alternatively, taking $A = h_\omega$, one could, for instance, construct the norm
\begin{align}
    \mathcal{C}_\Psi(t) \equiv \sqrt{\frac{1}{N}\, \sum_{\omega} \bigg|\diff{\langle h_\omega \rangle}{t} - \left\langle \diffp{h_\omega}{t} \right\rangle\bigg|^2}
    \label{eq:charge_error}
\end{align}
to quantify how well $\ket{\Psi}$ approximately solves the Schr\"odinger equation\----if $\ket{\Psi}$ solves the equation exactly, then the norm must vanish (since $[h_\omega,H]=0$). Note that this norm is evaluating an overall uncertainty of sorts, if we assume the uncertainty for each $h_\omega$'s constraint is uncorrelated to that of the rest.

One may attempt to further simplify matters by inserting the form of $h_\omega$ from Eq.~\eqref{eq:homega} into Eq.~\eqref{eq:ehrenfest}. Doing so, one obtains
\begin{widetext}
\begin{equation}
    - \diff{\langle J^z_\omega \rangle}{t} + 2 \, \diff{}{t} \left[ \mu \left\langle \sum_{\omega' (\neq \omega)} \frac{\vec{J}_\omega \cdot \vec{J}_{\omega'}}{\omega - \omega'} \right\rangle \right] = 2 \, \diff{\mu}{t} \, \left\langle \sum_{\omega' (\neq \omega)} \frac{\vec{J}_\omega \cdot \vec{J}_{\omega'}}{\omega - \omega'} \right\rangle.
\end{equation}
\end{widetext}
Using the chain rule on the left-hand side leads to a cancellation, leaving us with
\begin{equation}
    \diff{\langle J^z_\omega \rangle}{t} = 2 \mu \, \diff{}{t} \left\langle \sum_{\omega' (\neq \omega)} \frac{\vec{J}_\omega \cdot \vec{J}_{\omega'}}{\omega - \omega'} \right\rangle.
\end{equation}
Since $P_{z}(\omega) = 2 \langle J^z_\omega \rangle $, one may define the Ehrenfest error measure as
\begin{equation}
    \mathrm{Ehr}_\omega[\Psi(t)] \equiv \frac12 \, \diff{P_{z}(\omega)}{t} - 2 \mu \, \diff{}{t} \left\langle \sum_{\omega' (\neq \omega)} \frac{\vec{J}_\omega \cdot \vec{J}_{\omega'}}{\omega - \omega'} \right\rangle,
    \label{eq:ehror}
\end{equation}
where $\mathrm{Ehr}_\omega=0$ for each $\omega$ if $\ket{\Psi(t)}$ is the exact evolved state. 

Note that applying the Ehrenfest theorem again to the latter term of the above equation, and using Eq.\eqref{eq:homega} and the fact that $[h_\omega, H] = 0$, the above relation simplifies to
\begin{align}
    \frac12 \, \diff{P_{z}(\omega)}{t} &= \frac{1}{i} \left\langle [J^z_\omega, H] \right\rangle \nonumber\\
    &= 2\mu\hat{z}\cdot\braket{\sum_{\omega'(\neq\omega)}\vec{J}_{\omega'}\times\vec{J}_\omega},
\end{align}
which is simply the Ehrenfest theorem applied to $J^z_\omega$.

This condition could be used for a consistency check, to ensure that the many-body wave function obtained using any numerical approximation does indeed satisfy the Schr\"odinger equation to an acceptable level of accuracy. In particular, we propose the use of the maximum value of
\begin{align}
    \mathrm{max\:Ehr}[\Psi] \equiv \max_t \max_\omega \big|\mathrm{Ehr}_\omega[\Psi(t)]\big|
    \label{eq:maxehr}
\end{align}
as a measure to assess how accurately a particular time-evolution algorithm is calculating our many-body state. In the case of MPS calculations of real-time evolution, we point out that there may be a handful of initial time steps needed to transition an initial product state (where $D=1$ for all bonds) to an entangled state where virtual bonds have as many nontrivial singular values as the max bond dimension that we set for a given calculation. As such, we find that the algorithm requires a few time steps to ``warm up'' and accumulate enough singular values to well-approximate our desired wave function, and so Ehr may vary in a less well-behaved fashion until the state is evolved to $t\sim t_0+\omega_0^{-1}$; as a consequence, we choose the domain for evaluating $\mathrm{max\:Ehr}_\omega[\Psi(t)]$ to be $t>t_0+\omega_0^{-1}$.

\section{Results}
\label{sec:results}

We wish to assess the resources needed to implement TDVP methods for MPS calculations of the dynamics of our model. Our first step is to address how to determine the minimum bond dimension needed to accurately time evolve a many-body state describing a dense neutrino gas. We use the max Ehr quantity discussed in the previous section to do so. We find that max Ehr may tend to be larger for initial conditions that result in multiple spectral splits, even for methods calculating the entire wave function such as RK4. However, this error quantity turns out to be independent of $N$, according to calculations for $N<20$; as such, we propose the use of this quantity to determine when a MPS calculation is being carried out with insufficient precision. Using max Ehr provides a method to assess the accuracy at large values of $N$ for which comparison to the results of other numerical methods is unavailable.

Throughout this section we will consider two kinds of initial conditions for our many-body state: one where all neutrinos begin in the electron flavor state, $\ket{\nu_e}^{\otimes N}$, and one where the half of neutrinos with the lowest $\omega$ values start in the electron flavor state and the rest start in the $x$ flavor state, $\ket{\nu_e}^{\otimes N/2}\ket{\nu_x}^{\otimes N/2}$. As pointed out in, e.g., Refs.~\cite{Birol:2018qhx,Patwardhan:2021rej}, these conditions will result in one and two spectral splits, respectively. By taking these two cases into consideration, we may observe how our MPS methods handle evolved states with differing numbers of spectral splits. 

\begin{figure*}[htbp]
    \includegraphics[width=0.80\linewidth]{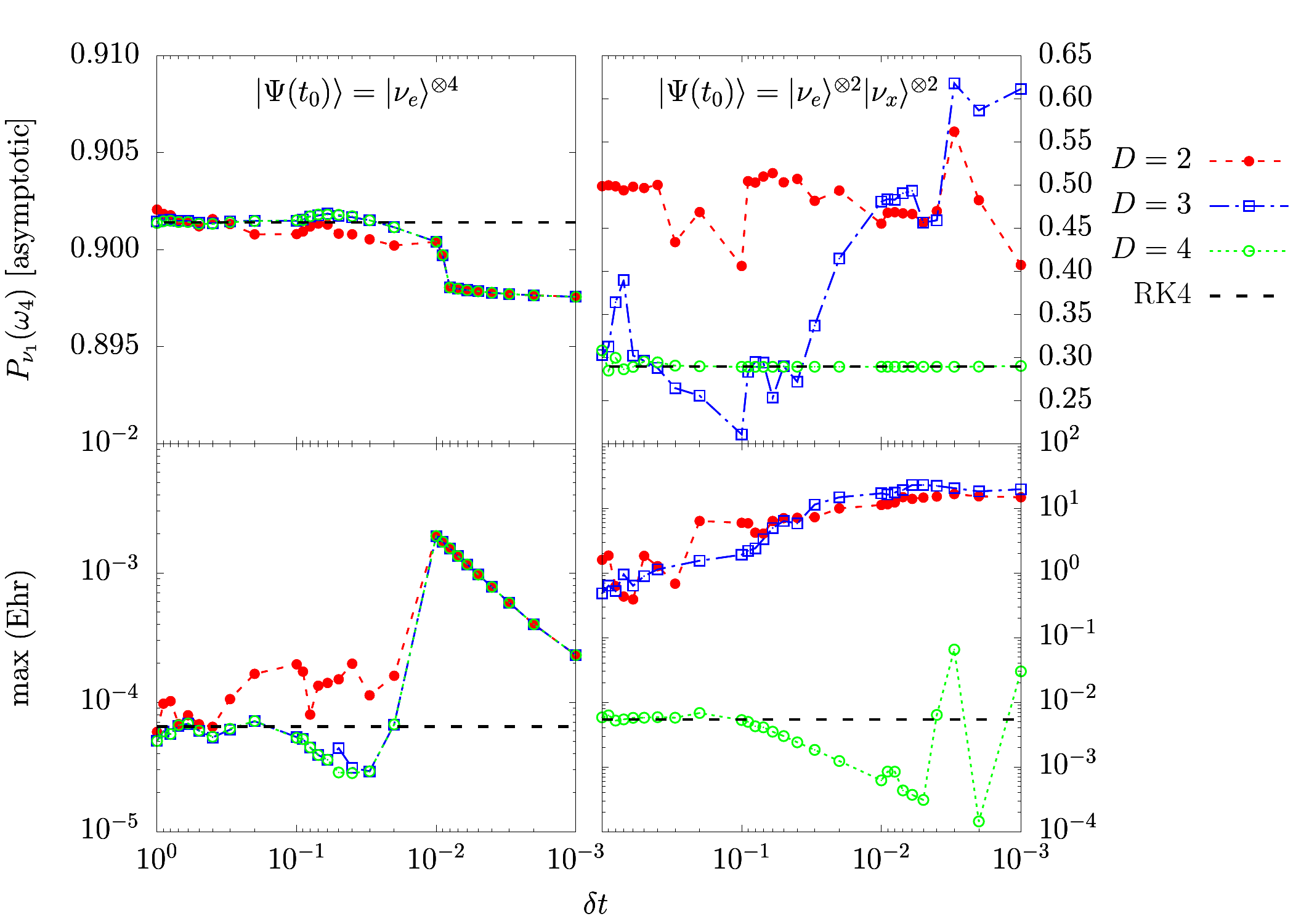}
    \caption{
    Determination of appropriate step size for tensor network calculation of $N=4$ with a box spectrum in $\omega$ for different initial conditions. 
    Top row: Asymptotic value ($t=1000\omega_0^{-1}$) of the probability $P_{\nu_1}(\omega_N)$ as defined by Eq.~\eqref{eq:polarprob} using tensor networks with different bond dimension cutoffs $D$ versus the magnitude of the time step $\delta t$ for two different initial conditions: left: single-flavor initial condition $\ket{\Psi(t_0)}=\ket{\nu_e}^{\otimes4}$; and right: mixed-flavor initial condition $\ket{\Psi(t_0)}=\ket{\nu_e}^{\otimes2}\ket{\nu_x}^{\otimes2}$. 
    Bottom row: Error value max Ehr, defined by Eq.~\eqref{eq:maxehr}, maximized over all times $t$ and $\omega$: left: $\ket{\Psi(t_0)}=\ket{\nu_e}^{\otimes4}$; and right: $\ket{\Psi(t_0)}=\ket{\nu_e}^{\otimes2}\ket{\nu_x}^{\otimes2}$. 
    We plot a flat, dashed line for the result from RK4 using the time-dependent step size outlined in Sec.~\ref{sec:methods-exact}, as a basis of comparison for the different MPS runs. 
    In the left case, we find that for larger choices of time-step sizes $\delta t$, there is some limited flexibility in choosing the maximum bond dimension $D$, where only $D=4$ consistently comes in close agreement with the results of earlier RK4 calculations of $P_{\nu_1}$. However, as $\omega_0 \delta t$ shrinks to $O(10^{-2})$, we find that the final data for the wave function deviate from the RK4 results and converge slowly to an incorrect value for $P_{\nu_1}(\omega_N)$. Correspondingly, we find that max Ehr is relatively well-behaved, with values $O(10^{-4})$ until $\delta t\lesssim10^{-2}$, at which point we find max Ehr to increase by orders of magnitude, for all choices of $D$. This increase in error for these calculations with a decrease in the choice of $\delta t$ is indicative of a growth in total truncation error from the end of each step in the GSE-TDVP2 algorithm. In the right case, we find that only the use of $D=4$ produces results from our MPS method that agree closely with those of RK4. In kind, we find that max Ehr is consistently orders of magnitude smaller for $D=4$ than for $D=3,4$. Though obscured by the greater discrepancies for $D<4$ for the latter initial condition, there is a larger discrepancy for step sizes $10^{-3}$ and $3\times10^{-3}$ correlating with the sudden increases in max Ehr.
    }
    \label{fig:N4}
\end{figure*}

We begin with a demonstration for the use of max Ehr in the relatively simple case of $N=4$ calculations using GSE-TDVP2~\cite{PhysRevB.102.094315}. Knowing the results of RK4 calculations to be very precise, we may compare the values of probability $P_{\nu_1}(\omega_N)$ defined in Eq.~\eqref{eq:polarprob} after evolving to $\mu(t)\ll\omega_0$ and the max Ehr for the wave function $\Psi(t)$ [over all times $t>t_0+\omega_0^{-1}$] with $D\geq2$, shown in Fig.~\ref{fig:N4}. 
In the cases of each initial condition, we find that there is eventually a growth in error as one tries to decrease the time-step size $\delta t$ to be too small. Even for the maximum physical bond dimension $D=2^{\lfloor N/2\rfloor}=4$, results converge to values of $\lim_{t\to\infty}P_{\nu_1}(\omega_N)$ that differ from those obtained using RK4. Recall that there are not only finite time-step errors related to forward-integrating with the effective one/two-body Schr\"odinger equations for each tensor in train, but also truncation and projection errors as outlined in Appendix~\ref{sec:tensor_net_form}. Since GSE introduces new singular values, which may not be physical if the number of values exceeds $2^{\lfloor N/2\rfloor}$, there will be a truncation error after each time step, independent of the size of the time step. Therefore, there is a truncation error that could grow at least linearly with the total number of time steps. (As a consequence there can be a nontrivial truncation error when using GSE even for $D=2^{\lfloor N/2\rfloor}$.) Additionally, comparing the two initial conditions, we observe that these max Ehr values are of a certain order of magnitude for certain choices of $D$ and $dt$ that match with max Ehr in RK4 results, suggesting that this quantity may help diagnose if our choices of parameters for evolution of our MPS can reasonably approximate the  exact many-body wave function. For the mixed-flavor initial condition, we find that there is a region of $\delta t$ values for each initial condition where max Ehr is small for at least certain values of $D$, while for other $D$ values always result in max Ehr that is orders of magnitude larger. For the single-flavor initial condition, there is less sensitivity overall to choice of $D$, but there is a shared trend of converging to the wrong results after $\delta t\lesssim0.01$, where max Ehr quickly trends upward.

\begin{figure*}[htbp]
    \includegraphics[width=0.9\linewidth]{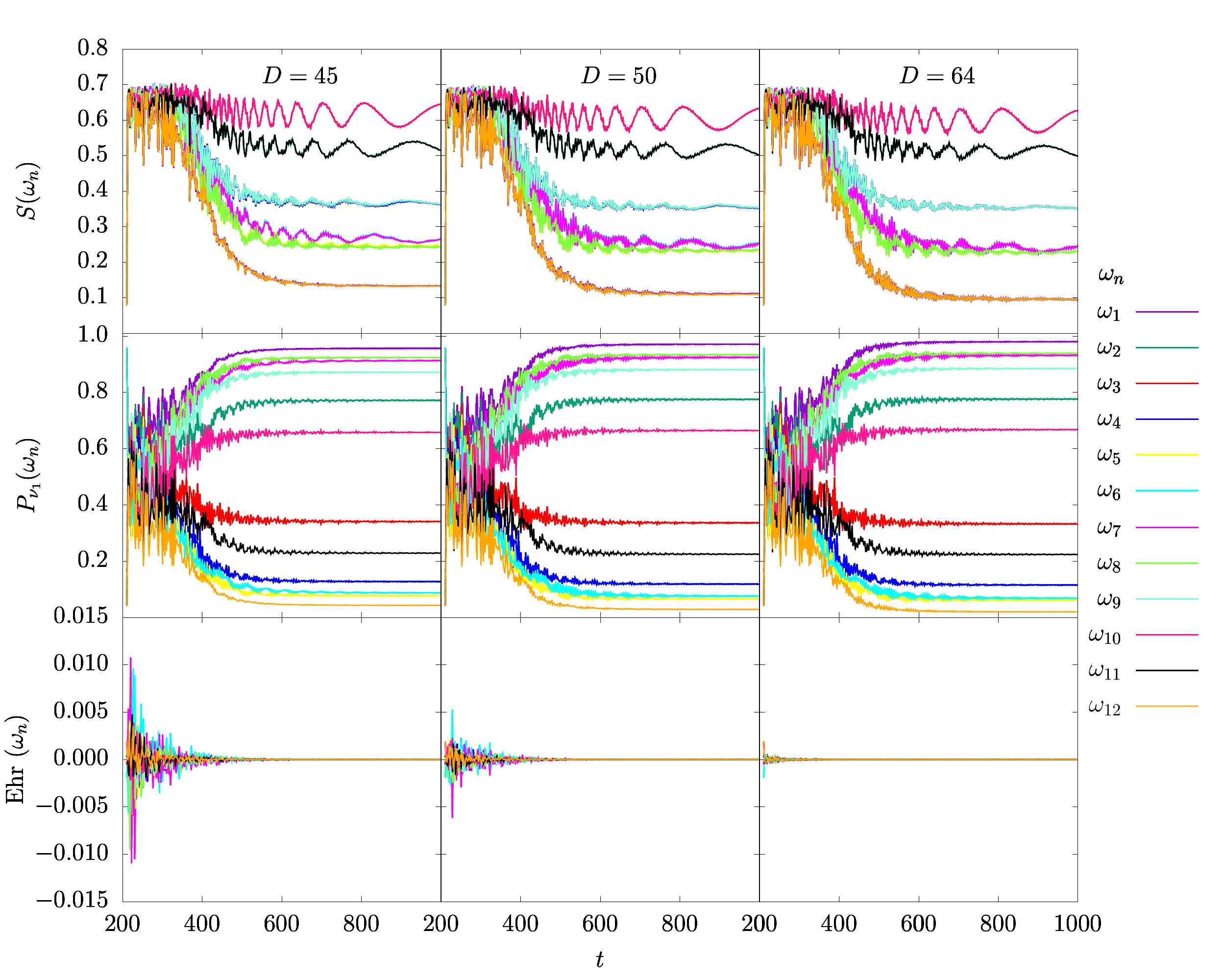}
    \caption{
    Time evolution from $\mu(t)=5\omega_0$ to $\mu\ll\omega_0$ of entanglement entropy (top row), probability $P_{\nu_1}$ (middle row), {and Ehrenfest error \lq\lq Ehr\rq\rq\ defined in Eq.~\eqref{eq:ehror}} (bottom row), for each neutrino mode in an ensemble of $N=12$ with an initial spectrum of $\ket{\nu_e}^{\otimes6}\ket{\nu_x}^{\otimes6}$, evolved using a time step of $\delta t=0.01$ and a maximum bond dimension of $D=45$ (left column), $50$ (middle column), or $64$ (right column). Note that $D=64=2^{\lfloor12/2\rfloor}$ is the largest physical choice of maximum bond dimension for a system of 12 neutrinos in two flavors, implying there is no error due to projection of the evolution equations; additionally, for $D>32$, the only bond for which singular values are being ignored are on the central virtual bond (i.e., between sites 6 and 7). Consequently, we find that Ehr is most well-behaved for this case, while Ehr can reach values that are an order of magnitude larger, particularly at early times in the evolution. Correspondingly, we can carefully observe that as we decrease $D$, there is an overestimation of $S(\omega)$, especially for neutrino modes with the lowest values of $S$. As per Eqs.~\eqref{eq:entropy} and~\eqref{eq:1dens-eigen}, $P_{\nu_1}$ is therefore bounded only to slightly smaller values. Interestingly, there is a greater discrepancy in results between $D=45$ and $50$ than that between $D=50$ and $64$, implying that our projection error while evolving this system grows dramatically as $D$ decreases beyond the point of $D\sim50$. 
    }
    \label{fig:N12}
\end{figure*}

\begin{figure}[htbp]
    \includegraphics[width=0.99\linewidth]{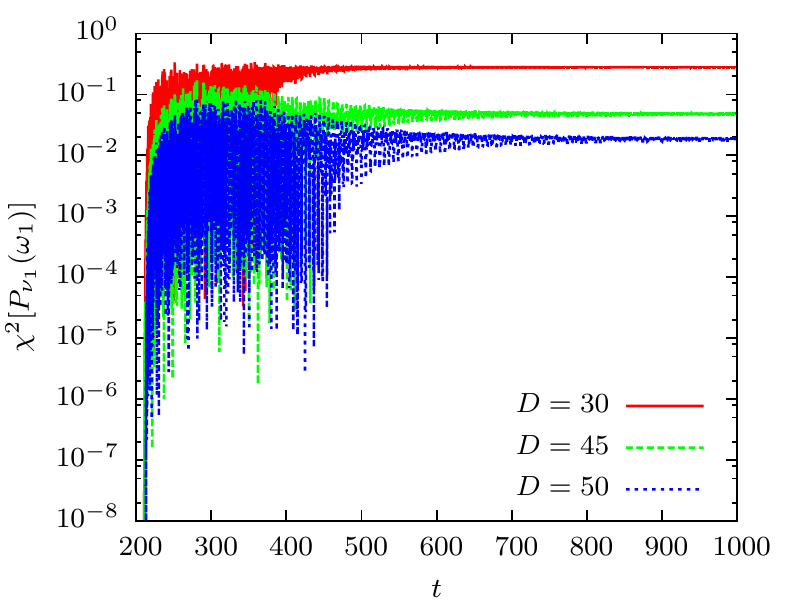}
    \caption{Discrepancy in mass-1 probabilities as a function of time, between bond dimension $D=30$, $45$, or $50$ and $D=64$, as measured by $\chi^2[P_{\nu_{1}}(\omega_{1})]=\big|P_{\nu_{1}}^2(\omega_{1})|_D-P_{\nu_{1}}^2(\omega_{1})|_{D=64}\big|/P_{\nu_{1}}^2(\omega_{1})|_{D=64}$ with respect to the maximum bond dimension $D=64$. As in Fig.~\ref{fig:N12}, we consider a system of $N=12$ evolving from the state $\ket{\nu_e}^{\otimes6}\ket{\nu_x}^{\otimes6}$ at $\mu(t)=5\omega_0$.
    As one excludes more virtual bond singular values by decreasing the value of $D$, again progressively greater discrepancies in the predictions of probabilities $P_{\nu_1}$ are found, due to progressively greater overestimates of the neutrino mode's entanglement entropy.}
    \label{fig:Pnu_N12_chi2}
\end{figure}

Taking this understanding of max Ehr, we can more easily approach calculations with larger $N$. For example, we can consider the case of $N=12$, testing our method with the mixed-flavor initial condition described earlier in this section. We find that a time-step size $\delta t=0.01\omega_0^{-1}$ is appropriate for varied choices; in particular, we depict in Fig.~\ref{fig:N12} the time evolution of $P_{\nu_1}$, $S$, and $\mathrm{Ehr}$ for each neutrino mode with the choices of bond dimension cutoff $D=45$, $50$, and $64$. We find that the differences in $P_{\nu_1}$ and $S$ for $D=64$ and $50$ are relatively modest, corresponding with a difference in max Ehr values that is slightly less than an order of magnitude. In comparison, there is a larger discrepancy between $D=50$ and $45$, both with max Ehr as well as $S$ and $P_{\nu_1}$. Specifically, one can see here that the lowest values of $S$ are---perhaps counterintuitively---overestimated by the use of too small of a cutoff $D$. (This observation is reflected also in 2TDVP calculations without the addition of GSE.) Consequently, the values of $P_{\nu_1}$ permitted by Eq.~\eqref{eq:entropy} are thus more tightly bound, resulting in another observable difference, whereby probabilities $P_{\nu_1}$ cannot approach $0$ and $1$. This relationship is isolated in Fig.~\ref{fig:Pnu_N12_chi2}, where we show, as an example, the evolution in the discrepancies for $P_{\nu_1}(\omega_1)$ as a function of time for $D=50$, $45$, and $30$ from the maximum $D=64$. Correspondingly, there is a yet larger growth in max Ehr between $D=50$ and $45$, suggesting a tipping point in one's choice of decreasing $D$ where our predictions become progressively less accurate.

\begin{figure}[htbp]
    \includegraphics[width=0.99\linewidth]{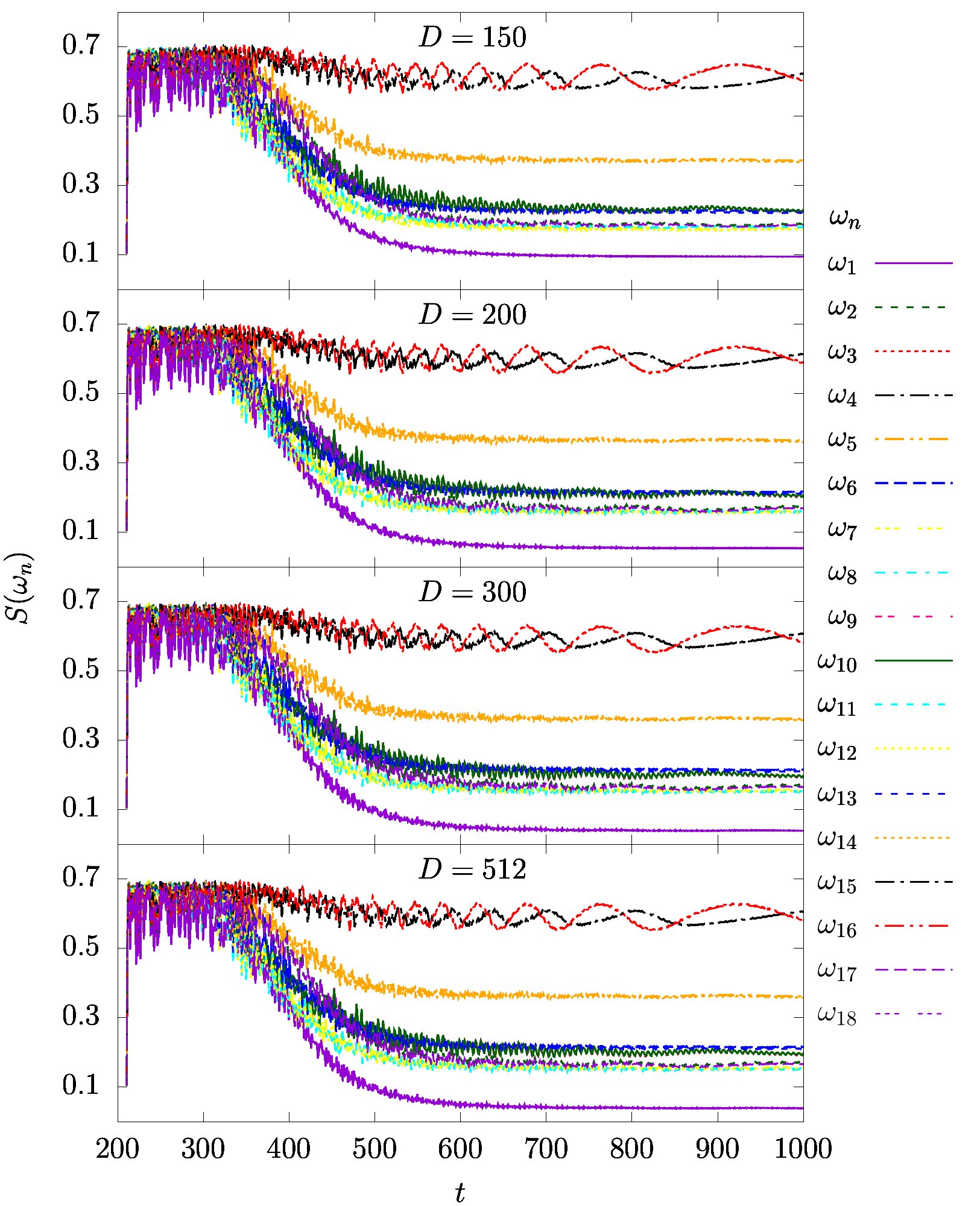}
    \caption{Entanglement entropy of each neutrino mode $\omega$ as a function of time for a system of $N=18$, evolving an initial state $\ket{\nu_e}^{\otimes 9}\ket{\nu_x}^{\otimes 9}$ from $\mu(t)=5\omega_0$, using $\delta t=0.01$. We show calculations using maximum bond dimension values of $D=150$, $200$, $300$, and $512$, respectively from top to bottom.
    In particular, one can notice that the neutrino modes with the lowest entanglement entropy values throughout most of the time evolution are also those whose values are most greatly overestimated as we decrease $D$.}
    \label{fig:N18_entropy}
\end{figure}

\begin{figure}[htbp]
    \includegraphics[width=0.99\linewidth]{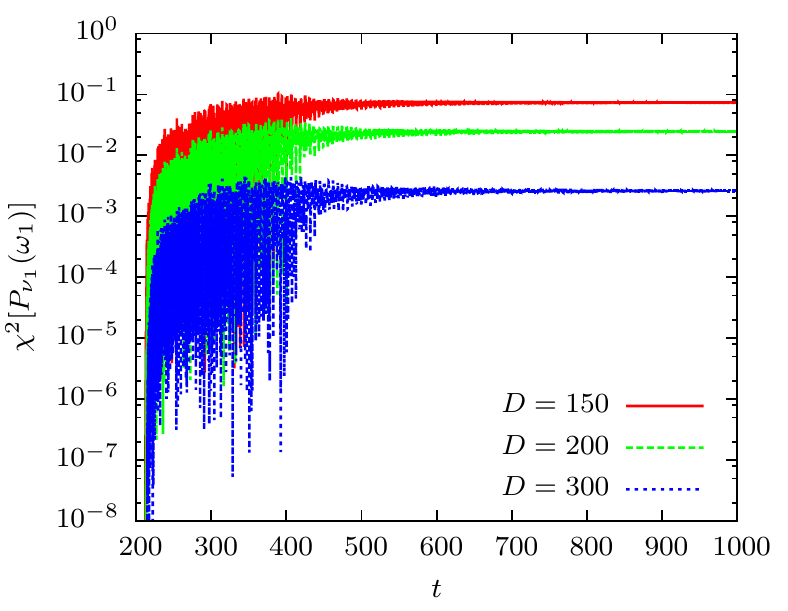}
    \caption{Same as Fig.~\ref{fig:Pnu_N12_chi2}, except for a system of $N=18$ evolving from $\ket{\nu_e}^{\otimes9}\ket{\nu_x}^{\otimes9}$, where the new basis for comparison is with the maximum bond dimension $D=512$, as in Fig.~\ref{fig:N18_entropy}. Again, neutrinos far from the spectral split frequencies experience greater overestimation of their entanglement entropy as the bond dimension cutoff $D$ is decreased; consequently, the discrepancy in probability $P_{\nu_1}$, as measured by $\chi^2$, grows. }
    \label{fig:Pnu_N18_chi2}
\end{figure}

\begin{figure}[htbp]
    \includegraphics[width=0.99\linewidth]{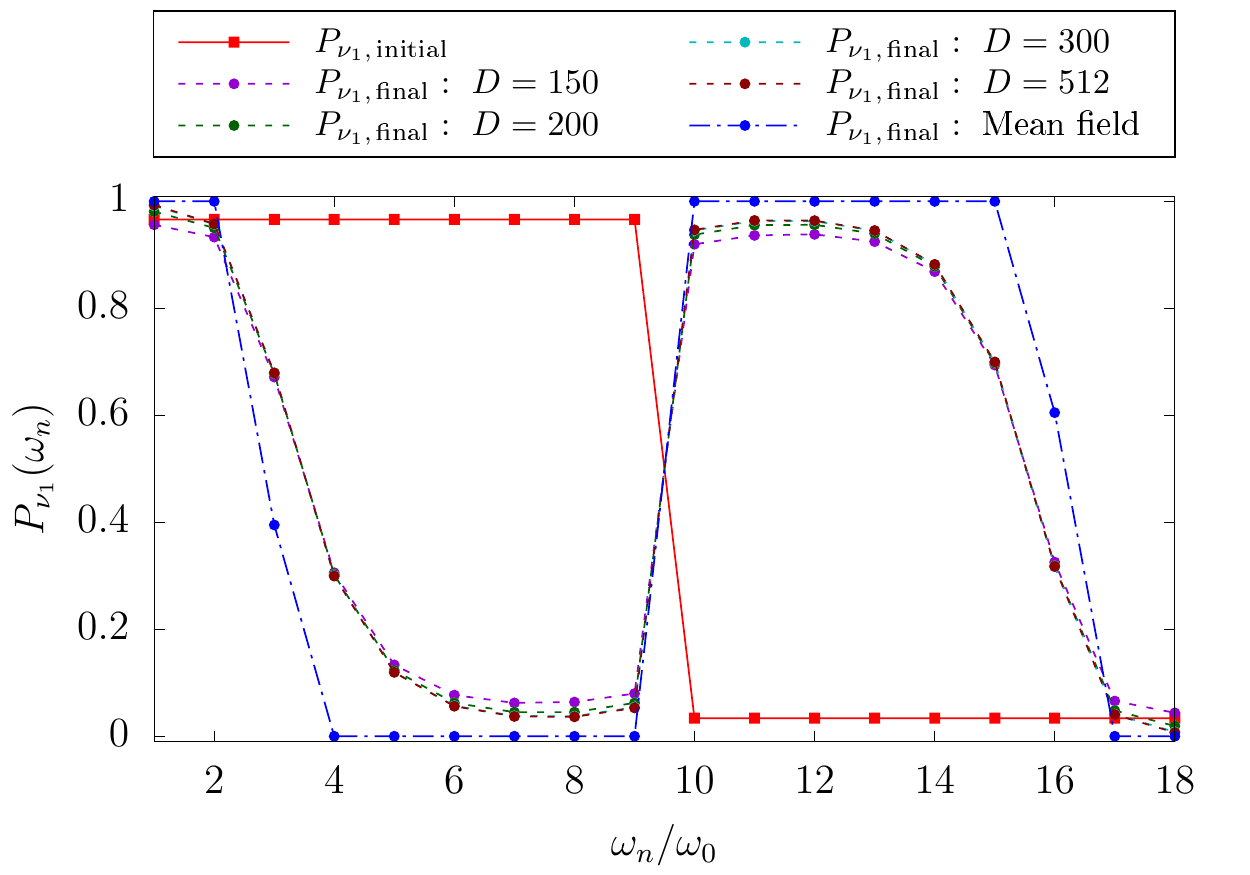}
    \caption{Final ($t=1000\omega_0^{-1}$) mass-1 probability spectra calculated via GSE-TDVP2, evolving the initial state $\ket{\nu_e}^{\otimes9}\ket{\nu_x}^{\otimes9}$ from $\mu(t_0)=5\omega_0$, with $\delta t=0.01$. As in Fig.~\ref{fig:N18_entropy}, we compare the results using bond dimensions $D=2^{\lfloor18/2\rfloor}=512$, $300$, $200$, and $150$.
    Interestingly, the locations as well as the widths of the spectral splits ($\omega/\omega_0\sim3,15$) are unchanged by the reduction in bond dimension.
    However, we find that the range of permitted values in probability $P_{\nu_1}$ is decreased due to the overestimation in entanglement entropy observed in Fig.~\ref{fig:N18_entropy}; therefore, the probabilities for modes furthest for the spectral splits are restricted to values closer to $1/2$.
    }
    \label{fig:N18_spectra}
\end{figure}

For these relatively small values of $N$, our system does not yet reflect a benefit in terms of complexity in our MPS calculations with growing $N$. However, we may look to cases of yet larger $N$ in order to check that the $D$ needed to obtain results at a desired level of accuracy (according to e.g., max Ehr) does not grow exponentially in $N$. 
In Fig.~\ref{fig:N18_entropy}, we again consider the predicted time evolution of entanglement entropy for each $\omega$, starting from the mixed-flavor initial condition but with $N=18$; here, the largest physical bond dimension would be $D=512$, which we compare against the choices of cutoff $D=300$, $200$, and $150$. 
As in the case of $N=12$, we find that lowering $D$ too far can result in overestimates of the lowest values of $S(\omega)$ at a given time. 
We depict an example of this effect in Fig.~\ref{fig:Pnu_N18_chi2}, where values of $P_{\nu_1}(\omega_1)|_{D}$ deviate further from $P_{\nu_1}(\omega_1)|_{D=512}$ throughout the evolution as $D$ is decreased. 
Furthermore, in Fig.~\ref{fig:N18_spectra}, we explore how these overestimates impact the prediction of the spectral split for this system; as in smaller $N$ calculations, we find that the location and width of the split are unaffected, while the permitted range of values for probability $P_{\nu_1}$ are more limited. 
However, in contrast with smaller $N$ calculations, we find that one can reasonably approximate our system using values of $D\lesssim300$. In the same vein, max Ehr values are $\sim10^{-3}$ for $D=150$ and $200$ and $\sim10^{-4}$ for $D=300$ and $512$. This result suggests that a shrinking fraction of the complete set of singular values of the MPS are required by the GSE-TDVP2 algorithm to obtain accurate results as we increase $N$. 

Carrying out an analogous comparison of time evolution results for varied $N$ with these initial conditions, we summarize the temporal complexity of using GSE-TDVP2 with the smallest necessary bond dimensions and largest time-step sizes to reproduce the desired level of accuracy in max Ehr when compared to the methods of RK4 and Lanczos propagation. We show these results as well as extrapolations from these data in Fig.~\ref{fig:time}. Extrapolated fit functions for RK4 and Lanczos propagation are based on $O(N^42^N)$ complexity discussed in Sec.~\ref{sec:methods-exact}, while there is an empirical trend line $e^{-aN^2+bN+c}$ with $a,b,c>0$ to depict one possible extrapolation of the data obtained from MPS methods for the initial condition resulting in one spectral split, as its well-behavedness permitted a greater number of calculations to display.\footnote{To fit other functions such as polynomials $N^a$ to MPS computation times obtained would require yet larger-$N$ calculations to discern what polynomial would be most appropriate in the regime of $N$ values for which we already see $D/2^{\lfloor N/2\rfloor}$ decrease with $N$.} In case of GSE-TDVP2, for an initial state $\ket{\nu_{e}}^{\otimes N}$, the computation time scales more favorably. Whereas, for a mixed initial state $\ket{\nu_{e}}^{\otimes N/2}\otimes\ket{\nu_{x}}^{\otimes N/2}$, scaling of computation time is less easily controlled, though calculations carried out on multiple cores (not shown in figure) suggest that choosing the smallest $D$ to reproduce max Ehr values comparable to those of RK4 and Lanczos results in a shrinking fraction $D/2^{\lfloor N/2\rfloor}$ as $N$ increases implying better scaling than RK4 and Lanczos scaling. Note that the data are obtained considering the smallest maximum bond dimension and lowest number of time steps required for a desired accuracy up to $N\lesssim20$. These parameters are expected to increase with $N$. Consequently, the scaling might differ from the ones presented here. Though bond dimension appears to grow more than linearly in $N$ for our system, the growth could still be polynomial. 
Consequently, we see that the growth in complexity with $N$ for our MPS methods appears optimistic, requiring a shrinking fraction $D/2^{\lfloor N/2\rfloor}$ as $N$ grows, in contrast with the manifestly exponential scaling of RK4 and Lanczos propagation. 

\section{Conclusions} 
\label{sec:conclusion}

This paper addresses the computational challenge of investigating the nature of collective neutrino oscillations in a dense neutrino gas. 
Employing the growing body of work in the MPS community to reformulate the problem in terms of a chain of effective two-body Schr\"odinger equations, we have investigated how TDVP methods (including GSE) may help in this line of inquiry. 
Furthermore, we have defined measures for error for many-body calculations from the instantaneously conserved charges of our Hamiltonian, to assess how well a calculated many-body wave function reflects the solution to our evolution equations. 
Where there is limited entanglement in our system (i.e., especially in conditions resulting in at most one spectral split), there are some use cases for GSE-TDVP2 in which MPS methods appear to scale much more favorably with $N$ than other numerical methods such as Runge-Kutta (RK4). 
However, existing TDVP methods scale much less favorably when we consider initial conditions that require $D\sim2^{\lfloor N/2\rfloor}$, in which case the temporal and spatial complexity will scale exponentially with $N$ anyhow---although this growth will be slower than in the cases of the other algorithms.

The methods of GSE-TDVP$n$ are new, and it remains to be seen whether further augmentations or improvements to existing methods can be made to permit yet larger $N$ calculations ($\sim100$) in a reasonable time frame. 
Conversely, one may point to several initial conditions in our model as systems in which quantum entanglement grows in such a way that even MPS representations may not efficiently treat the problem. In such cases there may still be an exciting opportunity for digital or analog quantum simulation hardware to assist in studying our Hamiltonian~\cite{Yeter-Aydeniz:2021olz}. 
As we ramp up to larger $N$ calculations with techniques such as those we have discussed, we hope to learn more about the scaling behavior of correlations in collective neutrino oscillations with larger systems.

\begin{acknowledgments}
We thank Frank Pollmann and Steven White for helpful correspondence and A.~Carosso and A.~Friedland 
for helpful conversations. This work was supported in part by the U.S.~Department of Energy, Office of Science, Office of High Energy Physics, under Awards No.~DE-SC0019465 and No.~DE-FG02-95ER40907. 
It was also supported in part by the U.S. National Science Foundation Grants No.~PHY-2020275 and No.~PHY-2108339. The work of A.~V.~P. was supported by the U.S.~Department of Energy under Contract No.~DE-AC02-76SF00515.
The work by S.~N.~C. was supported in part by Google~Asia~Pacific.
Tensor network calculations were performed using the {\small{ITensory}} Library (version 3.1.10)~\cite{itensor} and the {\small{TeNPy}} Library (version 0.9.0)~\cite{tenpy}.
\end{acknowledgments}

\appendix

\section{Details of Many-body Wave Function Calculations}
\label{sec:exact-details}

In this appendix, we elaborate on methods used to obtain the time-evolved wave function from Eq.~\eqref{eq:schrod} using the full $2^N$-dim Hilbert space. In particular, we provide greater depth on how to perform further explanation regarding the use of Lanczos propagation.

In our application, the many-body state is forward-integrated according to Eq.~\eqref{eq:schrod} by applying a time-evolution operator: 
$  \ket{\Psi(t+\delta t)} = U(t+\delta t;t)\ket{\Psi(t)}$.
Formally, one treats the time-evolution operator $U$ by a Magnus expansion:
\begin{align}
    U(t+\delta t;t) &= e^{\sum_{j=1}^\infty\Omega_j(t,\delta t)}; 
    \label{eq:magnusexp} \\ 
    \Omega_1(t,\delta t) &= -i\int_t^{t+\delta t}\mathrm{d}t'\:H(t'), 
    \label{eq:magnus1} \\ 
    \Omega_2(t,\delta t) &= -\frac{1}{2}\int_t^{t+\delta  t}\mathrm{d}t'\int_{t}^{t'}\mathrm{d}t''\:[H(t'),H(t'')], 
    \label{eq:magnus2} \\
    \Omega_3(t,\delta t) &= \frac{i}{6}\int_{t}^{t+\delta t}\mathrm{d}t'\int_{t}^{t'}\mathrm{d}t''\int_{t}^{t''}\mathrm{d}t'''
    \nonumber\\&\phantom{=\frac{i}{6}\int}\times\{[[H(t'),H(t'')],H(t''')]
    \nonumber\\&\phantom{=\frac{i}{6}\int\times\{}+[[H(t'''),H(t'')],H(t')]\},
    \label{eq:magnus3} 
\end{align}
and so on. After truncating the Magnus expansion, one can approximate the  time-evolution operator acting on the state $\ket{\Psi(t)}$ by implementing the Lanczos algorithm with full reorthogonalization. 

At the core of the Lanczos propagation method is the use of a low-dimensional effective basis in which one can apply an approximate time-evolution operator given by Eqs.~\eqref{eq:magnusexp}--\eqref{eq:magnus3}. 
{Because we found no noticeable advantage in going beyond the first term 
in the Magnus expansion, this task means computing the naive evolution operator 
$u= \exp( - i H(t) \delta t).$ 
Rather than exponentiating $H$ in the full basis, which would be prohibitive for large dimensions, 
one constructs iteratively a $k$-dimensional subspace, the Krylov subspace, by orthogonalizing 
the set of vectors $\{[H(t)]^\ell\ket{\Psi(t)}\}$ for $\ell = 0,\ldots, k-1$; the representation 
of $H(t)$ in this subspace is provided automatically as part of the algorithm. 
}
While in 
many applications the Lanczos algorithm is used to 
approximate extremal eigenpairs of a Hamiltonian~\cite{stewart2001matrix,RevModPhys.77.427},
here we use it to construct  the
time evolution operator  by approximate  spectral 
decomposition.

{ 
For instance, let $K$ be the $2^N\times k$ matrix, {constructed by the orthonormalized 
basis vectors of the Krylov subspace,} that maps from the full Hilbert space to the Krylov subspace, in which the Hamiltonian is tridiagonal, and let 
$V$ be a $k\times k$ real orthogonal matrix that diagonalizes $H(t)_K = K \, H(t) K^T$, the approximate Hamiltonian at time $t$ projected into the Krylov subspace. 
As $ \epsilon = V K H(t) K^T V^T $ is diagonal, it is trivial to exponentiate: 
$u = \exp(-i \,\epsilon \, \delta t)$.  Then one transforms back from the (approximate) 
eigenbasis to the Krylov basis and then finally to the original space,
\begin{align}
    \Psi(t+\delta t) \approx K^T \, V^T \, u(t+\delta t;t) \, V \, K \Psi(t),
\end{align}
which is the time evolution or Lanczos propagation by one time step~\cite{cremon2013adaptive}.
}

Notably, in the Lanczos propagation method, when taking the same time steps as those from our RK4 procedure, we find that the Lanczos algorithm needs very few iterations (i.e., typically 3--4) in order to arrive at a convergent result for the time-evolved wave function $\ket{\Psi(t+\delta t)}$ as the number of iterations of the algorithm is increased. Additionally, we find negligible difference in the results of the evolution whether or not the second term of the Magnus expansion $\Omega_2$ is included. Moreover, we find an agreement between this Lanczos propagation and RK4, whereby expansion beyond order $(\delta t)^4$ is unnecessary to replicate results obtained exactly with the Bethe ansatz method.{\footnote{ In close analogy with the Lanczos method, we could also propose the Fer expansion (see, e.g., \cite{Wilcox:1967zz}) as another numerical technique that approximates Eq.~\eqref{eq:time-evo-op-step}. 
Due to the extreme similarity at order $(\delta t)^4$ to RK4, we find that computation times are very similar between the two methods. Consequently, results produced from this method are not shown. 
}}

In the case of a constant Hamiltonian [i.e., constant $\mu(t)$], the Lanczos algorithm would greatly simplify the complexity of the time evolution that results from applying the many-body Hamiltonian to an initial state $\ket{\Psi(t)}$, as the bulk of the time steps could be taken via calculations within the scalably small Krylov subspace. However, the case of a time-dependent Hamiltonian does not share this benefit, as the eigenbasis is evolving as well as the wave function, implying that we must change our basis out of the Krylov subspace following each time step. As a consequence, the time-dependent Lanczos propagation still suffers from the same difficulty as did RK4, in which a Hamiltonian that grows exponentially in $N$ must be applied to a wave function repeatedly to time evolve our ensemble.

\section{Tensor Network Formalism}
\label{sec:tensor_net_form}

In this Appendix, we provide greater detail into the MPS treatment of time evolution for long-range interacting systems such as that described by our Hamiltonian in this paper. First, we will elaborate on how we obtain different forms of MPS descriptions for our many-body wave function. Once this procedure has been outlined in greater detail, we will expound upon how the TDVP algorithm can be performed.

When we perform the truncation in the singular values with a bond dimension cutoff $D$ to a state such as that in Eq.~\eqref{eq:left-canon}, where each bond dimension $D_j$ is replaced with $\min\{D_j,D\}$, we must normalize our state by imposing on the remaining entries of the tensor train the following constraint:
\begin{align}
    \sum_{\alpha_j=e,x}\psi_L^{\alpha_j}(j)^\dagger \psi_L^{\alpha_j}(j) = \mathbb{1}_{D_j\times D_j},
    \label{eq:left-norm}
\end{align}
which we refer to as ``left-normalization.'' This constraint also fixes a ``gauge,'' the transformation of which leaves the MPS form unchanged (i.e., under insertions of $G_{j}G^{-1}_{j}$ between each bond $1\leq j<N$, where $G$ is a $D_{j}\times D_{j}$ invertible matrix). For Eq.~\eqref{eq:left-norm}, we define $D_N\equiv1$. 
We can repeat this same process of Schmidt decomposition while instead starting from the rightmost indices, and we replace the tensor symbols $\psi_L\mapsto \psi_R$ 
and the bond indices $\beta_j\mapsto\bar{\beta}_{j+1}$ and dimensions $D_j\mapsto\bar{D}_{j+1}$ 
to denote this change in method, yielding a so-called ``right-canonical'' form. After performing the same truncation in the bond dimensions, we impose the ``right-normalization'':
\begin{align}
    \sum_{\alpha_j=e,x}\psi_R^{\alpha_j}(j) \psi_R^{\alpha_j}(j)^\dagger = \mathbb{1}_{\bar{D}_j\times\bar{D}_j},
    \label{eq:right-norm}
\end{align}
where we define $\bar{D}_1\equiv1$. 
Going forward, it will also be useful to define $D_0,\bar{D}_{N+1}\equiv1$ to include the cases of $j=1,N$ automatically. 

We can then use these two decompositions to write left and right blocks of the MPS wave function: 
\begin{align}
    \ket{\Phi_{L,\beta_j}(1\!:\!j)} &\equiv \sum_{\alpha_1,\ldots,\alpha_{j}=e,x}[\psi_L^{\alpha_1}(1)\cdots \psi_L^{\alpha_{j}}(j)]_{\beta_{j}}
    \nonumber \\
    &\phantom{\equiv\sum_{\alpha_{j},\ldots,\alpha_N=e,x}[\psi_L^{\alpha_1}(1)} \times \ket{\nu_{\alpha_1},\ldots,\nu_{\alpha_{j}}}, \\
    \ket{\Phi_{R,\bar{\beta}_j}(j\!:\!N)} &\equiv \sum_{\alpha_{j},\ldots,\alpha_N=e,x}[\psi_R^{\alpha_{j}}(j)\cdots \psi_R^{\alpha_N}(N)]_{\bar{\beta}_{j}} \nonumber \\
    &\phantom{\equiv\sum_{\alpha_{j},\ldots,\alpha_N=e,x}[[\psi_L^{\alpha_1}(j)}  \times\ket{\nu_{\alpha_{j}},\ldots,\nu_{\alpha_N}}.
\end{align}
with which we construct the ``mixed-canonical'' form\footnote{To include the cases of $j=1,N$ automatically, we take $\ket{\Phi_L(j\!:\!k)}$ and $\ket{\Phi_R(j\!:\!k)}$ for $j>k$ to be trivial factors.}
\begin{align}
    \ket{\Psi} =& \sum_{\alpha_j=e,x} \sum_{\beta_{j-1}=1}^{D_{j-1}} \sum_{\bar{\beta}_{j+1}=1}^{\bar{D}_{j+1}} [\psi_{C}^{\alpha_j}(j)]_{\beta_{j-1}\bar{\beta}_{j+1}} 
    \nonumber \\
    &\times \ket{\Phi_{L,\beta_{j-1}}(1\!:\!j-1)}\ket{\nu_{\alpha_j}}\ket{\Phi_{R,\bar{\beta}_{j+1}}(j+1\!:\!N)},
    \label{eq:one-site_train}
\end{align}
where for the center site $j$ we have the $D_{j-1}\times\bar{D}_{j+1}$ matrix $\psi^{\alpha_j}_C(j)=\psi^{\alpha_j}_L(j)C(j)=C(j-1)\psi^{\alpha_j}_R(j)$ with a $D_{j}\times\bar{D}_{j+1}$ matrix $C(j)$ containing the singular values for the ``virtual bond'' between sites $j$ and $j+1$.
Moreover, using these definitions of left and right blocks, we can define orthonormal projection operators 
\begin{align}
    P_L(1\!:\!j) &\equiv \sum_{\beta_j=1}^{D_j}\ket{\Phi_{L,\beta_j}(1\!:\!j)}\!\bra{\Phi_{L,\beta_j}(1\!:\!j)}, \\
    P_R(j\!:\!N) &\equiv \sum_{\bar{\beta}_j=1}^{\bar{D}_{j}}\ket{\Phi_{R,\bar{\beta}_j}(j\!:\!N)}\!\bra{\Phi_{R,\bar{\beta}_j}(j\!:\!N)}.
\end{align}

The particular MPS form in Eq.~\eqref{eq:one-site_train} is immediately useful in the 1TDVP. As described earlier in Sec.~\ref{sec:methods-mps-tdvp}, the 1TDVP algorithm involves approximating Eq.~\eqref{eq:schrod} by
\begin{equation}
    i\diff{}{t}\ket{\Psi(t)} = P_{T_\Psi M}H(t)\ket{\Psi(t)},
\end{equation}
where the projection operator onto the tangent space, $P_{T_\Psi M}$, is given by\footnote{Here, we use a notational convention whereby Kronecker product (left-~or right-)multiplication by $P_L(j\!:\!k)$ or $P_R(j\!:\!k)$ for $j>k$ is defined to be the trivial operation of multiplication by the scalar value 1, as opposed to a nontrivial tensor product.} 
\begin{align}
    P_{T_\Psi^{(1)} M} =& \sum_{j=1}^N P_{L}(1\!:\!j-1) \otimes \mathbb{1}_{2\times2} \otimes P_{R}(j+1\!:\!N) 
    \nonumber\\
    &- \sum_{j=1}^{N-1} P_{L}(1\!:\!j)\otimes P_{R}(j+1\!:\!N)
\end{align}
in the 1TDVP method; here, the $j$th site is the one ``active'' site in a given step, evolving exactly according to 
\begin{equation}
    i\diff{}{t}\psi_C(j) = H(j)\psi_C(j),
    \label{eq:1site-evo}
\end{equation}
where $H(j)$ is an effective one-site Hamiltonian at site $j$ obtained using the projection operators described above. 
Notably, this equation does not permit changes in bond dimension between sites and therefore limits the growth of entanglement in the system as well if the initial state is, for example, a simple product state (i.e., $D=1$).

In order to observe growth of entanglement as the many-body state evolves, we require a generalization of our earlier tensor train decompositions into left, right, and center blocks where we permit the center block to include multiple sites $j,\ldots,k$: $\psi_C(j\!:\!k)$ such that we can write the entire state as 
\begin{align}
    \ket{\Psi} = \sum_{\alpha_j,\ldots,\alpha_k=e,x} \sum_{\beta_{j-1}=1}^{D_{j-1}} \sum_{\bar{\beta}_{k+1}=1}^{\bar{D}_{k+1}} [\psi_{C}^{\alpha_j\cdots\alpha_k}(j\!:\!k)]_{\beta_{j-1}\bar{\beta}_{k+1}} 
    \nonumber \\
    \times \ket{\Phi_{L,\beta_{j-1}}(1\!:\!j-1)}\ket{\nu_{\alpha_j}\cdots\nu_{\alpha_k}}\ket{\Phi_{R,\bar{\beta}_{k+1}}(k+1\!:\!N)}.
    \label{eq:multi-site_train}
\end{align}
We provide diagrammatic forms for presenting a two-site center tensor as well as left- and right-normalized one-site tensors that can be chained together by contraction over virtual bonds to obtain a complete MPS of a wave function, using a style in keeping with the diagrammatic conventions presented in Ref.~\cite{Haegeman_2016}. 
\begin{figure*}[htbp]
\begin{center}
    \subfloat[\label{fig:2site}]{
	    \includegraphics[width=0.32\textwidth]{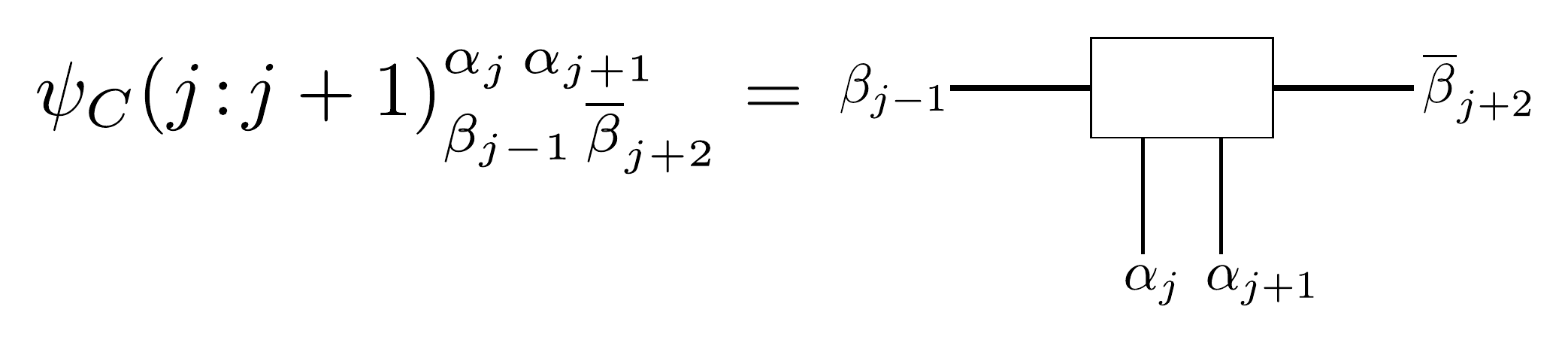}}
 	~
 	\subfloat[\label{fig:leftsite}]{
 	    \includegraphics[width=0.32\textwidth]{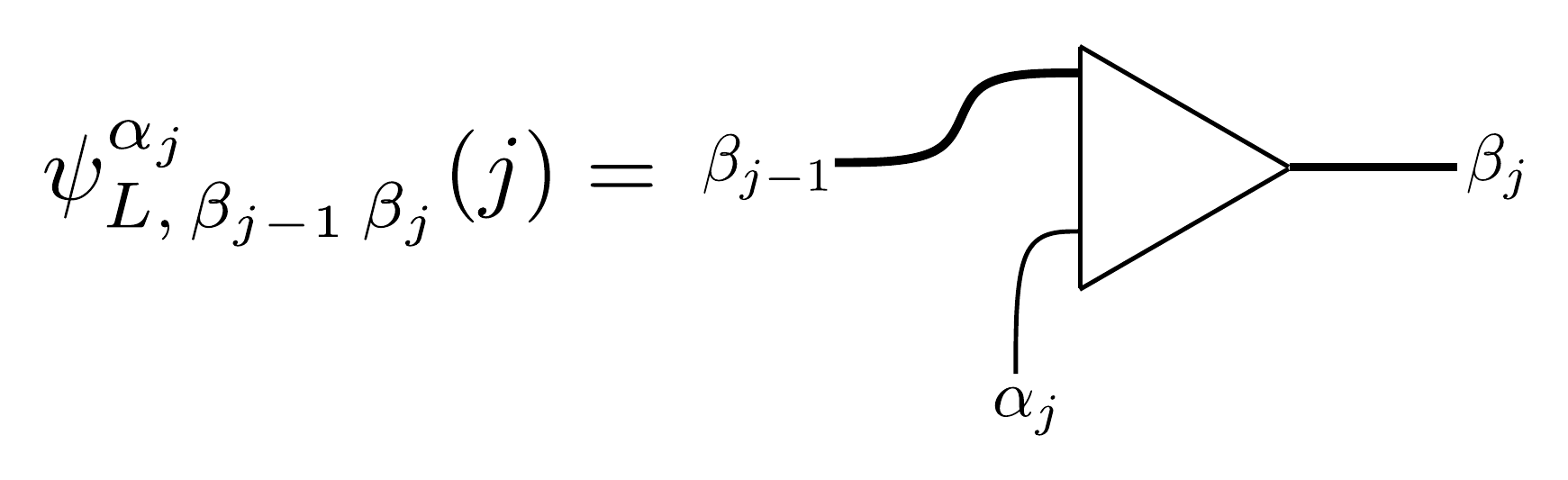}}
	~
	\subfloat[\label{fig:rightsite}]{
	    \includegraphics[width=0.32\textwidth]{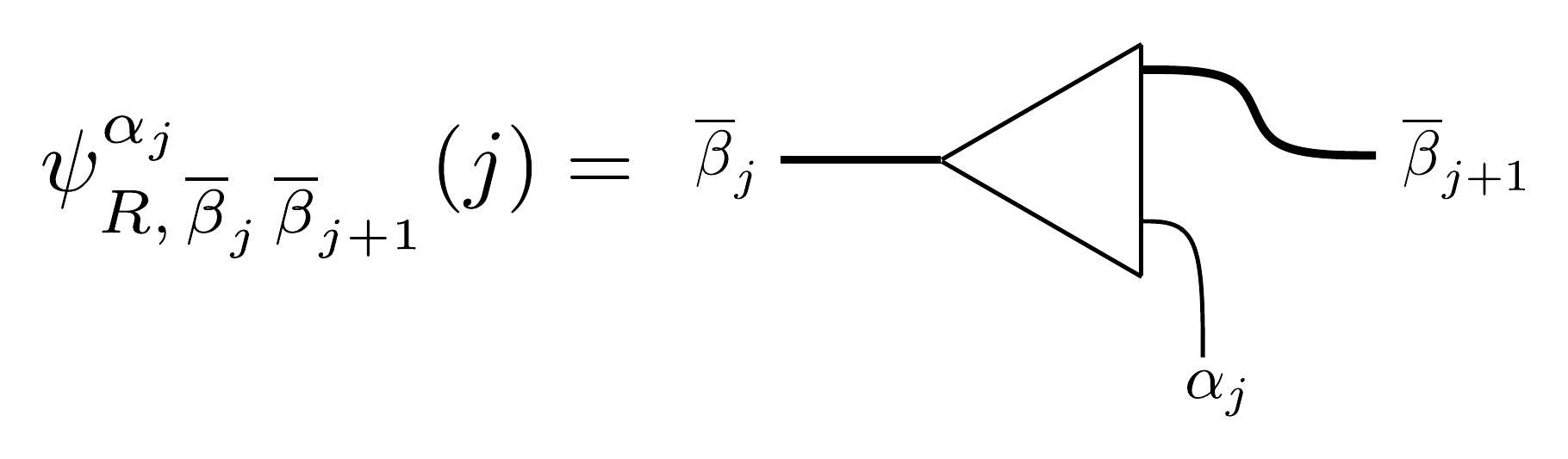}}
    
\end{center}
    \caption{The tensor elements serving as the basic building blocks of tensor trains used to represent states and operators as MPSs and MPOs, respectively. In (a) we depict a tensor for a pair of center sites $j$ and $j+1$ where $1\leq j<N$. Symbols $\beta$ and $\bar{\beta}$ denote the indices for wide legs by which the factors of the train are connected via tensor multiplication, while symbols $\alpha$ denote the indices for thin legs by which the factors of the train connect to particular basis kets or bras for the corresponding sites. In (b) and (c) we depict a left- and right-normalized tensor for a site $j$, respectively. In all subfigures, we use $\beta$ and $\bar{\beta}$ to denote internal indices for bonds to left- and right-normalized tensors, respectively, while we use the convention that bond indices $\beta_0,\beta_N,\bar{\beta}_1,\bar{\beta}_{N+1}\equiv1$ are entirely, as there is no connection to a further site for the case of an open boundary condition. 
    In keeping with the style of Ref.~\cite{Haegeman_2016}, we also use equilateral triangles pointing rightward (leftward) to symbolize a left- (right-)normalized tensor and rectangles to symbolize center site tensors, written in symbolic form by Eqs.~\eqref{eq:left-norm} and~\eqref{eq:right-norm}.  
    A contraction over a given virtual bond index $\beta_j$ or $\bar{\beta}_j$ involves a sum over index values $1,\ldots,D_j$ or $1,\ldots,\bar{D}_j$, respectively, while a contraction over an external flavor/mass index $\alpha_j$ involves a sum over $e,x$ or masses $1,2$. 
    } 
    \label{fig:2tdvp_tensors}
\end{figure*}

\begin{figure*}[htbp]
\begin{center}
    \subfloat[\label{fig:2site_psi0}]{
        \includegraphics[width=0.50\textwidth]{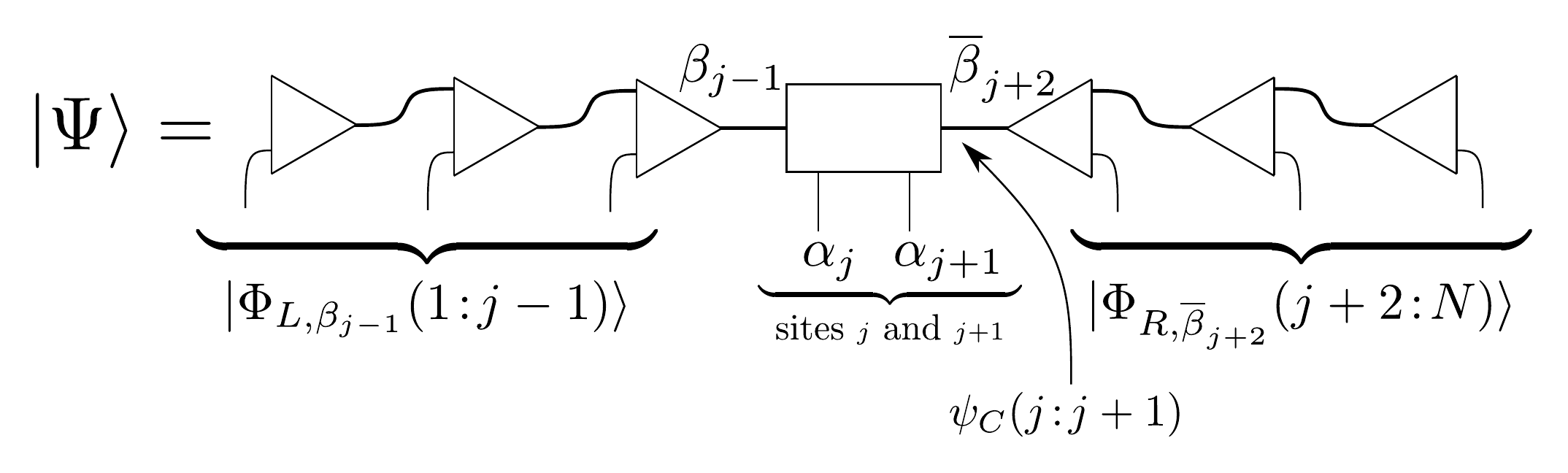}}
    \\
    \subfloat[\label{fig:2site_H}]{
        \includegraphics[width=0.50\textwidth]{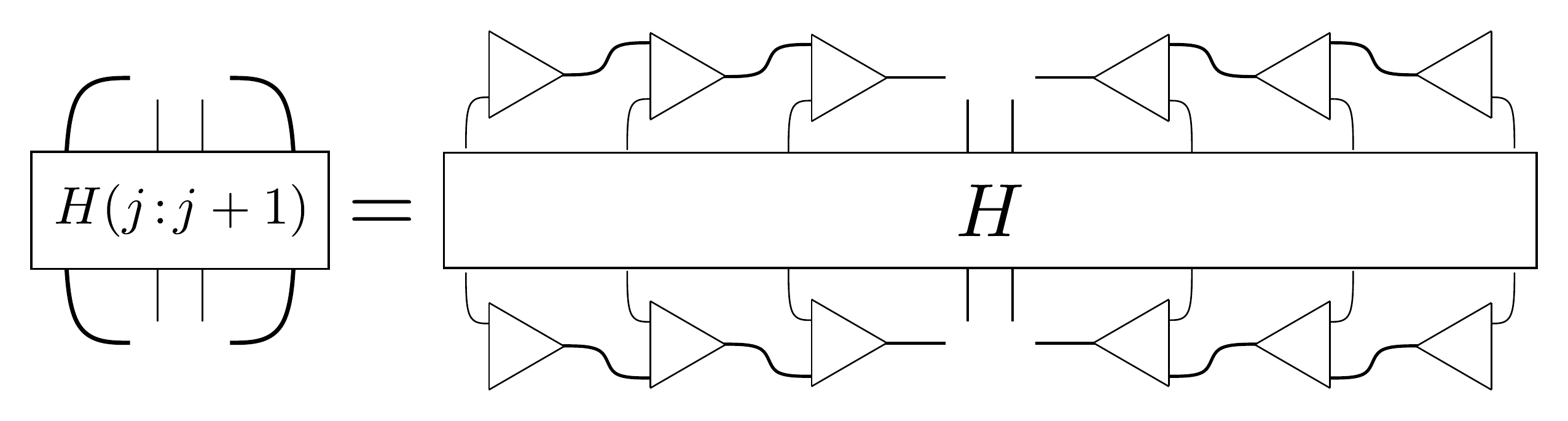}}
    \\
    \subfloat[\label{fig:2site_Hpsi0}]{
        \includegraphics[width=0.75\textwidth]{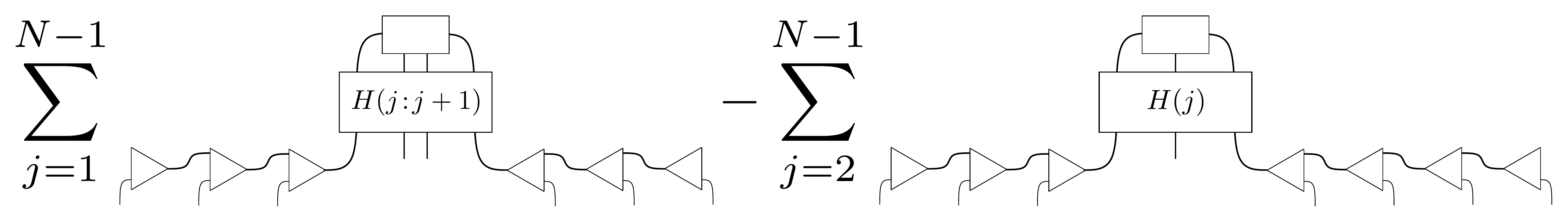}}
\end{center}
    \caption{
    The tensor network diagrams representing (a) a MPS in mixed-canonical form, (b) one of the two-site effective Hamiltonians in 2TDVP, and (c) the resulting MPS after applying two-site effective Hamiltonians at each pair of sites and subtracting the MPS and subtracting the MPS resulting from applying the one-site effective Hamiltonians at each nonedge site.
    In~(b) and~(c), we allow $1\leq j<N$. In equation form,~(a) is written out specifically in Eq.~\eqref{eq:multi-site_train}, while (b) depicts the result of applying a two-site projector term from Eq.~\eqref{eq:two-site_proj} to the complete Hamiltonian, $H$, from Eq.~\eqref{eq:saham}. Finally,~(c) depicts the result of combining~(a) and~(b) at each site and subtracting the analogous results of applying one-site projectors in the same method instead of the two-site projectors; this quantity represents the right-hand side of Eq.~\eqref{eq:2tdvp}. 
    All conventions of these diagrams are defined in Fig.~\ref{fig:2tdvp_tensors}. 
    } 
    \label{fig:2tdvp_formalism}
\end{figure*}

For example, one can depict a MPS with a two-site center by Fig.~\ref{fig:2tdvp_tensors}(a). We are then prepared to define a tangent space projector for the case of two active sites:
\begin{align}
    P_{T_\Psi^{(2)} M} =& \sum_{j=1}^{N-1} P_{L}(1\!:\!j-1) \otimes \mathbb{1}_{4\times4} \otimes P_{R}(j+2\!:\!N) 
    \nonumber\\
    & - \sum_{j=2}^{N-1} P_{L}(1\!:\!j-1) \otimes \mathbb{1}_{2\times2} \otimes P_{R}(j+1\!:\!N).
    \label{eq:two-site_proj}
\end{align}
We can then define an effective multisite Hamiltonian by applying projection operators such as the first series of terms above to $H(t)$ from the left [e.g., depicted in Fig.~\ref{fig:2tdvp_formalism}(b) for a two-site center]. Replacing the choice of $P_{T_\Psi M}$ with the two-site projection operator defined here, we obtain the two-site TDVP (2TDVP) equations where
\newlength{\myl}
  \settowidth{\myl}{$\times$}
\begin{widetext}
\begin{align}
    P_{T^{(2)}_\Psi M} H(t)\ket{\Psi(t)} = \sum_{j=1}^{N-1} \sum_{\alpha_j,\alpha_{j+1}=e,x}\sum_{\beta_{j-1}=1}^{D_{j-1}}\sum_{\bar{\beta}_{j+2}=1}^{\bar{D}_{j+2}}
    &[H(j\!:\!j+1)\psi_C(j\!:\!j+1)]_{\beta_{j-1}\bar{\beta}_{j+2}}^{\alpha_j\alpha_{j+1}} \nonumber \\
    &\phantom{[H(j)\psi_C(j)]_{\beta_{j-1}\bar{\beta}_{j+1}}^{\alpha_j}} \hspace{-\myl}\!\! \times \ket{\Phi_{L,\beta_{j-1}}(1\!:\!j-1)}\ket{\nu_{\alpha_j}\nu_{\alpha_{j+1}}}\ket{\Phi_{R,\bar{\beta}_{j+2}}(j+2\!:\!N)}
    \nonumber \\ 
    - \sum_{j=2}^{N-1} \sum_{\alpha_j=e,x}\sum_{\beta_{j-1}=1}^{D_{j-1}}\sum_{\bar{\beta}_{j+1}=1}^{\bar{D}_{j+1}}
    &[H(j)\psi_C(j)]_{\beta_{j-1}\bar{\beta}_{j+1}}^{\alpha_j}\ket{\Phi_{L,\beta_{j-1}}(1\!:\!j-1)}\ket{\nu_{\alpha_j}}\ket{\Phi_{R,\bar{\beta}_{j+1}}(j+1\!:\!N)}.
    \label{eq:2tdvp}
\end{align}
\end{widetext}
We visually summarize the steps of 2TDVP in Fig.~\ref{fig:2tdvp_formalism}, again using a style in keeping with that of Ref.~\cite{Haegeman_2016} for 1TDVP. 
Forward-integrating the many-body state $\ket{\Psi(t)}$ according to the Schr\"odinger equation with the above form for the right-hand side is the 2TDVP algorithm, whose full list of instructions is given in Ref.~\cite{Haegeman_2016}. More compactly, in order to carry out a step of 2TDVP, one forward-integrates the effective two-site evolution equations 
\begin{equation}
    i\diff{}{t}\psi_C(j\!:\!j+1) = H(j\!:\!j+1)\psi_C(j\!:\!j+1),
    \label{eq:2site-evo}
\end{equation}
and subtracts from the resulting wave function the MPS obtained by backward-integrating the effective one-site equation in Eq.~\eqref{eq:1site-evo}. Pictorially, the left terms of Fig.~\ref{fig:2tdvp_formalism}(c) represent the expression in Eq.~\eqref{eq:2tdvp}, while each term's center sites [without contraction with the left- and right-canonical tensors represent Eq.~\eqref{eq:2site-evo}].

Additionally, the Tensor Network Python ({\small{TeNPy}}) library~\cite{tenpy} provides several functions to help set up a program that evolves a many-body state in a MPS representation via a time-dependent Schr\"odinger equation via TDVP. As each time step is performed in the MPS formalism via an application of the Lanczos algorithm, this procedure can be thought of as a tensor network analog of the Lanczos evolution performed with a complete many-body state in sparse matrix representation. The finite time-step error of this algorithm is of order $(\delta t)^3$, though the use of two active sites in our tangent space projections necessitates a truncation during singular value decomposition (SVD) to reduce the time-evolved $\psi_C(j\!:\!j+1)\mapsto\psi(j)\psi(j+1)$ that introduces error of size constant with respect to the choice of time-step size. {Also, in contrast with the unitary evolution of 1TDVP, normalization of the wave function is no longer automatically preserved with 2TDVP if a truncation is performed at the end of a time step; in this case, one may need to divide the MPS by its norm between steps.} As a consequence, it is (perhaps counterintuitively) desirable for the sake of precision to keep the time step from being taken as very small if one is to use 2TDVP over many time steps{, as suggested by Ref.~\cite{PAECKEL2019167998}}. 

Before we conclude this section, let us discuss a more recent augmentation to the TDVP algorithms involving another avenue for growth in bond dimension between time steps. In particular, we follow the GSE method proposed by Ref.~\cite{PhysRevB.102.094315} to be used prior to each time step of TDVP; notably, this method does not depend in principle upon the number of active sites used during $n$TDVP (where $n$ is the number of active sites), so this method introduces an algorithm for each choice of $n$: GSE-TDVP$n$. 

The first of two steps of the GSE is to gather the Krylov subspace by which we will extend the MPS $\ket{\Psi(t)}$. We can obtain $k-1$ states to extend the bond basis of $\Psi$ in a numerically stable fashion by replacing $[H(t)]^\ell$ with $[1-i\delta tH(t)]^\ell$, as a first-order expansion of $U(t+\delta t;t)\approx e^{-i\delta tH(t)}\approx 1-i\delta tH(t)$ for sufficiently small $\delta t$ produces smaller changes to the norm of our vectors, yielding:
\begin{align}
    \mathcal{K}_k(t,\delta t) &\equiv \mathrm{span}\{\ket{\Psi^{(\ell)}(t,\delta t)}\}_{\ell=0}^{k-1}
    \label{eq:mps-krylov} \\
    \mathrm{where}\quad \ket{\Psi^{(\ell)}(t,\delta t)} &\equiv [1-i\delta tH(t)]^\ell\ket{\Psi(t)}.
\end{align}
Empirically, one finds in using GSE-TDVP that only a small value of $k\sim5$ and relatively little accuracy in obtaining the extra Krylov states $\ell>0$ are typically needed, implying a much larger truncation parameter for SVD can be utilized in this step---let us call it $\varepsilon_K$---than that for representing our time-evolved wave function.

Now, let us introduce the second step of GSE, in which we use the basis of the Krylov subspace above to extend our MPS for $\ket{\Psi}$.\footnote{For the remainder of this explanation, we shall suppress the notation for time dependence, for brevity.} Let us start with a given left-canonical form for $\ket{\Psi}$ as in Eq.~\eqref{eq:left-canon} and $k-1$ additional basis MPSs $\ket{\Psi^{(\ell)}}$ ($\ell>0$) obtained in Eq.~\eqref{eq:mps-krylov}. The general goal of this step is to incorporate singular values from the Krylov basis as we rewrite $\ket{\Psi}$ as a MPS in right-canonical form via $N-1$ steps of SVD as mentioned earlier in this section. Starting from $j=N$ and working our way to $j=1$, we perform SVD at each site $j$ as the orthogonality center of $\ket{\Psi}$: $\psi_C(j)\mapsto C(j-1)\psi_R(j)$ and define a projection operator $P_R(j)\equiv1-\psi_R(j)^\dagger\psi_R(j)$. Then, to ensure the additional bond bases of our final MPS are orthogonal to those of the original MPS, we project the tensors at the orthogonality center for each $\Psi^{(\ell)}$: $\psi^{(\ell)}_C(j)\mapsto P_R(j)\psi^{(\ell)}_C(j)P_R(j)\equiv\psi_{C0}^{(\ell)}(j)$ and perform SVD on $\bigoplus_{\ell=1}^{k-1}\psi_{C0}^{(\ell)}(j)\mapsto \tilde{C}(j-1)\tilde{\psi}_R(j)$. (Note that any truncation parameter that we may use here $\varepsilon_M$ in mixing the Krylov states while neglecting small singular values needs neither to be the same as $\varepsilon_K$ from the earlier GSE step of obtaining the Krylov subspace nor to correspond to the truncation error of whatever $D_\mathrm{max}$ we use during the TDVP steps.) Finally, we use the resulting right-orthonormal tensor $\tilde{\psi}_R(j)$ to extend $\psi_R(j)$ via $\psi_R(j)\oplus\tilde{\psi}_R(j)$.

Extending our time-evolved state in this fashion between TDVP time steps has proved useful in the case of a GSE-TDVP1 calculation of the real-time evolution of the one-axis twisting model~\cite{PhysRevB.102.094315}. More generally, it was proposed that GSE-TDVP1 allows the user to enlarge the tangent space of the MPS manifold before each time step, thus permitting growth in bond dimension even in models involving various kinds of non-neighbor interactions; even without using two active sites during TDVP, the extra Krylov states allow calculations to grow the bond dimension over time. This augmentation of the TDVP method may come with additional benefits, such as permitting smaller bounds on bond dimension under certain circumstances and use of relatively large time steps compared to other methods. 

\bibliography{references}

\end{document}